\documentclass[a4paper,fleqn]{cas-dc}

\usepackage[numbers]{natbib}

\usepackage{hyperref}
\usepackage{natbib}
\usepackage{enumerate}
\usepackage{booktabs}
\usepackage{todonotes}
\usepackage{soul, color}
\usepackage{makecell}
\usepackage{multirow}
\usepackage{adjustbox}
\usepackage{pdflscape}
\usepackage{afterpage}
\usepackage{caption}
\usepackage{paralist}
\usepackage{tikz}
\usepackage[graphicx]{realboxes}
\usepackage{microtype}

\hypersetup{
  breaklinks   = true, 
  colorlinks   = true,    
  urlcolor     = blue,    
  linkcolor    = blue,    
  citecolor    = gray      
}

\soulregister\cite7
\soulregister\ref7
\soulregister\pageref7

\def\tsc#1{\csdef{#1}{\textsc{\lowercase{#1}}\xspace}}
\tsc{WGM}
\tsc{QE}
\tsc{EP}
\tsc{PMS}
\tsc{BEC}
\tsc{DE}

\begin{document}
\let\WriteBookmarks\relax
\def\floatpagepagefraction{1}
\def\textpagefraction{.001}
\shorttitle{A Taxonomy of Assets for the Development of Software-Intensive Products and Services}
\shortauthors{Zabardast, Gonzalez-Huerta, Gorschek, et~al.}

\title [mode = title]{A Taxonomy of Assets for the Development of Software-Intensive Products and Services}                  



\author[1]{Ehsan Zabardast}[
                        orcid=0000-0002-1729-5154]
\cormark[1]
\ead{ehsan.zabardast@bth.se}
\ead[url]{https://ehsanzabardast.com/}


\address[1]{Software Engineering Research Lab SERL,
            Blekinge Institute of Technology,
            Campus Karlskrona, Valhallav{\"a}gen 1, 
            Karlskrona, Sweden }

\author[1]{Javier Gonzalez-Huerta}[orcid=0000-0003-1350-7030]
\ead{javier.gonzalez.huerta@bth.se}
\ead[url]{http://gonzalez-huerta.net/}

\author[1, 2]{Tony Gorschek}
\ead{tony.gorschek@bth.se}
\ead[url]{https://gorschek.com/}
\address[2]{fortiss GmbH, Guerickestraße 25, 80805 Munich, Germany}

\author[1]{Darja \v{S}mite}[orcid=0000-0003-1744-3118]
\ead{darja.smite@bth.se}

\author[1]{Emil Al\'egroth}
\ead{emil.alegroth@bth.se}

\author[1,3]{Fabian Fagerholm}[orcid=0000-0002-7298-3021]
\ead{fabian.fagerholm@aalto.fi}
\address[3]{Department of Computer Science,
            Aalto University,
            Espoo, Finland}


\cortext[cor1]{Corresponding author}

\nonumnote{This research was supported by the KK foundation through the SHADE KK-H\"{o}g project under grant 2017/0176 and Research Profile project SERT under grant 2018/010  at Blekinge Institute of Technology, SERL Sweden.
}

\begin{abstract}
\noindent \textbf{Context:} Developing software-intensive products or services usually involves a plethora of software artefacts. Assets are artefacts intended to be used more than once and have value for organisations; examples include test cases, code, requirements, and documentation. During the development process, assets might degrade, affecting the effectiveness and efficiency of the development process. Therefore, assets are an investment that requires continuous management. 

\noindent Identifying assets is the first step for their effective management. However, there is a lack of awareness of what assets and types of assets are common in software-developing organisations. Most types of assets are understudied, and their state of quality and how they degrade over time have not been well-understood. 

\noindent \textbf{Method:} We perform a systematic literature review and a field study at five companies to study and identify assets to fill the gap in research. The results were analysed qualitatively and summarised in a taxonomy. 

\noindent \textbf{Results:} We create the first comprehensive, structured, yet extendable taxonomy of assets, containing 57 types of assets. 

\noindent \textbf{Conclusions:} The taxonomy serves as a foundation for identifying assets that are relevant for an organisation and enables the study of asset management and asset degradation concepts.
\end{abstract}




\maketitle

\section{Introduction}\label{sec:introduction}

The fast pace of the development of software-intensive products or services impacts the decision-making process both for design and operational decisions. Such products and services are engineered by ``applying well-understood practices in an organised way to evolve a product [or service] containing a nontrivial software component from inception to market, within cost, time, and other constraints.~\cite{klotins2018software}''

Acting fast to cope with change can compromise the values of the delivered product, environment, development process, and the assets involved, such as source code, test cases, and documentation~\cite{broy2018logical,zabardast2022assets}. 

Assets are software artefacts that are intended to be used more than once during the inception, development, delivery, or evolution of a software-intensive product or service~\cite{zabardast2022assets}. Assets can lose their value and degrade ``due to intentional or unintentional decisions caused by technical or non-technical manipulation of an asset or its associated assets during all stages of the product life-cycle.~\cite{zabardast2022assets}''

During the development of software-intensive products or services, decisions are usually made quickly to respond to change, focusing on fast delivery, leaving considerations on the assets involved as a secondary objective. Thus, these decisions often result in long-term negative consequences not only on the quality of the delivered product or service but also on the assets such as code, architecture, and documentation, to name a few.

Researchers and practitioners have traditionally used the technical debt (TD) metaphor~\cite{cunningham1992wycash} to refer to the impact of the intentional and unintentional sub-optimal decisions on assets as a consequence of meeting strict deadlines~\cite{kruchten2019managing,zabardast2022assets}. TD, in practice, has become a concept used to assess and reason about one asset in isolation, i.e., it considers the impact of sub-optimal decisions related to a specific asset~\cite{zabardast2022assets}. However, assets can degrade due to the propagation of degradation coming from other assets. For example, taking shortcuts in requirements elicitation (i.e., requirements TD), created under time pressure, lead to sub-optimal design and development decisions~\cite{mendez2016naming, zabardast2022assets}.

The TD metaphor does not consider the propagation of degradation on assets, it mainly focuses on code-based assets, and often it does not take non-technical assets such as documentation, environmental, and other supporting items into account~\cite{avgeriou2016managing,brown2010managingTD,rios2018tertiary}. This is evident from, for example, the tools available for measuring TD that tend to all focus on measuring code-related TD, as pointed out in the tertiary study by Rios et al.~\cite{rios2018tertiary}. Since the original TD metaphor was proposed in 1992, Software engineering has evolved to take a wider, more encompassing perspective on software development, including aspects such as processes and environments but also softer factors. However, this perspective is still fairly new, and it is unclear how assets shall be dealt with and what role they play in this context~\cite{zabardast2022assets}.

The motivation behind the work presented in this paper comes from industrial collaborations where the concept of TD did not fully cover their needs since code-based assets, although important, only represented a part of the challenge. From a practitioner's perspective, it is valuable to have the ability to identify what assets are available, which of them are important, and how and why they degrade. This is especially important as a product or service evolves over time, when assets are reused many times, and when they inevitably become subject to change and degradation. The degradation of assets is central to their maintenance and evolution~\cite{avgeriou2016managing}.

In our previous work~\cite{zabardast2022assets}, we have defined \emph{assets} and \emph{asset degradation}. We have articulated the importance of identifying and studying them. From a research perspective, a potential benefit of introducing \emph{assets} as a complementary concept to TD, with a broader definition, is that it takes all so-called ``items of value''~\cite{fernandez2019artefacts} into account and widens the view of what can hold value in the context of developing software-intensive products or services~\cite{zabardast2022assets}. The widened definition also fosters an understanding of what assets can be negatively impacted by degradation and, thereby, the understanding that overlooking said assets can be detrimental to the development and evolution of such products and services. However, to gain such understanding, we must first understand what assets are subject to the concept, warranting the need for the taxonomy of assets presented in this work. 

The paper is structured as follows: Section~\ref{sec:background_relatedwork} provides the background and related work on the topic. Section~\ref{sec:research_overview} describes the research methodology, which separately covers the literature review and the field study. The results are presented in Section~\ref{sec:results}, together with the proposed asset management taxonomy. Section~\ref{sec:discussion} discusses the principal findings and the implications of the results. The threats to validity are also discussed and addressed in Section~\ref{sec:discussion}. And finally, Section~\ref{sec:conclusions_futurework} presents the conclusion and the continuation of the work together with the future directions.

\section{Background and Related Work}\label{sec:background_relatedwork}

Assets related to software products or services have been studied previously. For instance, from a managerial perspective where the term asset is used to discuss how products in product lines are developed from a set of core assets with built-in variation mechanisms~\cite{northrop2007framework}, or making emphasis only on business- or market-related assets, like the in the works by Ampatzoglou et~al.~\cite{ampatzoglou2018reusability}, Cicchetti et~al.~\cite{cicchetti2016towards}, Wohlin et~al.~\cite{wohlin2016supporting}, and Wolfram et~al.~\cite{wolfram2020recognising}. In contrast, in this work, our focus is on the inception, development, evolution, and maintenance of software-related assets.

\subsection{Artefacts and Assets in Software Engineering}\label{sec:sub_related_artefacts}

Describing how a software system is envisioned, built, and maintained is part of the Software Development Processes (SDP)~\cite{sommerville2015software}. The SDP prescribes the set of activities and roles to manipulate software artefacts, e.g., source code, documentation, reports, and others~\cite{broy2018logical}. 

Artefacts in software engineering field have been traditionally defined as i) ``documentation of the results of development steps''~\cite{broy2018logical}; ii) ``a work product that is produced, modified, or used by a sequence of tasks that have value to a role''~\cite{fernandez2019artefacts}; and iii) ``a self-contained work result, having a context-specific purpose and constitutes a physical representation, a syntactic structure and a semantic content of said purpose, forming three levels of perception''~\cite{fernandez2019artefacts}. Software artefacts are, therefore, self-contained documentation and work products that are produced, modified or used by a sequence of tasks that have value to a role~\cite{fernandez2019artefacts}.

Understanding software artefacts, how they are structured, and how they relate to each other has a significant influence on how organisations develop software~\cite{broy2018logical}. The documentation in large-scale systems can grow exponentially; therefore, there is a need for structurally organising software artefacts~\cite{broy2018logical}.
Artefacts defined by most of the SDPs are monolithic and unstructured~\cite{tilley2002documenting}. The content of poorly structured artefacts is difficult to reuse, and the evolution of such monolithic artefacts is cumbersome~\cite{silvasoftware}. Therefore, different SDPs present various models for presenting software artefacts, e.g., the Rational Unified Process (RUP)~\cite{kroll2003rational,kruchten2000rational}. There are ways to classify and structure software artefacts based on well-known modelling concepts. Examples of such models are the work of Broy~\cite{broy2018logical} and Silva et~al.~\cite{silvasoftware}. Moreover, there are ontologies and meta-models to classify artefacts in specific software development areas (e.g., Idowu et~al.~\cite{idowu2022AMinML} Mendez et~al.~\cite{fernandez2010meta}, Zhao et~al.~\cite{zhao2009ontology}, and Constantopoulos and Doerr~\cite{constantopoulos1995component}).

However, the definitions of artefacts presented in the literature do not distinguish between:  

\begin{enumerate}[i]
    \item \textbf{Artefacts that have an inherent value for the development organisation} (i.e., an asset~\cite{zabardast2022assets}) from the artefacts that do not have any value for the organisation\footnote{Mendez et~al.~\cite{fernandez2019artefacts} define artefacts as having value for a role which is substantially different from having value for the organisation.}~\cite{zabardast2022assets}. The value of each asset is a property that can characterise its degradation, i.e., if an asset degrades, although it continues being an asset, its value for the organisation has degraded.

    \item \textbf{Artefacts that are intended to be used more than once} ``An asset is a software artefact that is intended to be used more than once during the inception, development, delivery, or evolution of a software-intensive product or service...~\cite{zabardast2022assets}''  ``If an artefact is used once and discarded or disregarded afterwards, it does not qualify as an asset.~\cite{zabardast2022assets}''. We believe that the following example clarifies the distinction between \textit{Temporary Artefacts} vs \textit{Assets}: An API description used by developers as a reference has value in the development effort. If changes are made (new decisions, new ways to adhere to components, etc.), but the API description is not updated to reflect this, the utility (value) of the API description becomes lower (the asset degrades). On the other hand, an automatically generated test result is not seen as an asset, as it is transient or intermediate. It is generally created as a work product used once to be ``transformed'' into ``change requests'', ``tickets'' or other management artefacts. Once transformed into the new asset ``change requests'', it is discarded. New test results reports will be created (and discarded) on each execution of the tests. Therefore, artefacts that are intended to be used more than once and have value for the organisation (i.e., assets) need to be monitored by the organisation since their continuous maintenance and evolution renders the need for exercising quality control.~\cite{reussner2019managed,zabardast2022assets}.

\end{enumerate}

\subsection{Technical Debt and Asset Degradation}\label{sec:sub_related_td}

Cunningham introduced the TD metaphor in 1992 to describe the compromises resulting from sub-optimal decisions to achieve short-term benefits~\cite{cunningham1992wycash}. All assets (which are per definition artefacts, too) are subject to TD, i.e., while assets are created, changed, and updated, they might incur TD~\cite{li2015systematic}.

The TD metaphor has been extended and studied by many researchers~\cite{rios2018tertiary}. It has been an interesting topic for both academia and industry, and it has grown from a metaphor to a practice~\cite{kruchten2012technical}. TD is currently recognised as one of the critical issues in the software development industry~\cite{besker2017time}. TD is ``pervasive'', and it includes all aspects of software development, signifying its importance both in the industry and academia~\cite{kruchten2019managing}. The activities that are performed to prevent, identify, monitor, measure, prioritise, and repay TD are called Technical Debt Management (TDM)~\cite{avgeriou2016managing,griffith2014simulation} and include such activities as, for example, identifying TD items in the code, visualising the evolution of TD, evaluating source code state, and calculating TD principal~\cite{rios2018tertiary}.

As the TD metaphor was extended to include different aspects of software development, various TD types were introduced~\cite{avgeriou2016managing}, e.g., requirements debt, test debt, and documentation debt. The introduction of different types of TD has led researchers to attempt to classify the different types and categories of TD. One of the earliest classifications of TD is the work of Tom et~al.~\cite{tom2012consolidated}. Other secondary and tertiary studies have been performed to summarise the current state of TD and TD types, e.g., by Lenarduzzi et~al.~\cite{lenarduzzi2019technical}, Rios et~al.~\cite{rios2018tertiary}, and Li et~al.~\cite{li2015systematic}.

The metaphor has been mainly \textit{operationalised} by researchers and practitioners on source code, i.e., code, design, and architecture~\cite{avgeriou2016managing}. The organisational, environmental and social aspects of TD have not received the same amount of attention~\cite{rios2018tertiary}. Moreover, TD types are often studied in isolation, not considering how one TD can be contagious to other assets and the permeation to other TD types~\cite{avgeriou2016managing}. For example, when the TD in code grows, a valid question would be whether and how it impacts the TD in tests and the extent of such impact.

Moreover, TDM activities have been investigated in several secondary studies with different perspectives, some focusing on tools, others on strategies, but there is still a lack of unified analysis aligning these different perspectives~\cite{rios2018tertiary}. 

Lastly, TD is generally studied in the current state of the software, and it is understudied with regards to the evolutionary aspects of TD, i.e., studying TD on a ``snapshot'' of a system is not enough~\cite{fernandez2017identification,power2013understanding}. Therefore, a more appropriate approach to studying TD is to study its evolution~\cite{brown2010managing}. It is only by periodically monitoring TD that we can study the economic consequences of TD~\cite{falessi2013practical}, determine the performance of the system in the future~\cite{fernandez2014guiding}, and create methods and frameworks to react quickly to the accumulation of TD~\cite{letouzey2012managing}.

In our previous work~\cite{zabardast2022assets}, we coined another concept \textit{Asset Degradation} ``as the loss of value that an asset suffers due to intentional or unintentional decisions caused by technical or non-technical manipulation of the asset, or associated assets, during all stages of the product life-cycle''~\cite{zabardast2022assets}. All assets can degrade, which will affect their value for the organisation in different ways. Degradation can be deliberate, unintentional, and entropy~\cite{zabardast2022assets}. \textit{Deliberate degradation} is introduced by taking a conscious decision, understanding its long-term consequences and accommodating short-term needs. \textit{Unintentional degradation} is introduced by taking a sub-optimal decision either because we are not aware of the other, better alternatives or because we cannot predict the consequences of selecting ``our'' alternative. Finally, \textit{entropy} is introduced just by the ever-growing size and complexity that occurs when software systems are evolving~\cite{Lehman1996,lehman1979understanding}. The degradation that is due to the continuous evolution of the software, and which is not coming directly from the manipulation of the asset by the developers, is entropy~\cite{zabardast2022assets}.

\subsection{Taxonomies in Software Engineering}\label{sec:sub_related_taxonomy}

Scientists and researchers have long used taxonomies as a tool to communicate knowledge. Early examples are noted in the eighteen century, for instance, the work of Carl von Linn\'e~\cite{von1735systema}. Taxonomies are mainly created and used to communicate knowledge, provide a common vocabulary, and help structure and advance knowledge in a field~\cite{glass1995contemporary,kwasnik1992role,usman2017taxonomies}. Taxonomies can be developed in one of two approaches; top-down, also referred to as enumerative, and bottom-up, also referred to as analytico-synthetic~\cite{broughton2015essential}. The taxonomies that are created using the top-down method use the existing knowledge structures and categories with established definitions. In contrast, the taxonomies that use the bottom-up approach are created using the available data, such as experts’ knowledge and literature, enabling them to enrich the existing taxonomies by adding new categories and classifications~\cite{unterkalmsteiner2014taxonomy}.

Software engineering (SE) is continually evolving and becoming one of the principal fields of study with many sub-areas. Therefore, the researchers of the field are required to create and update the taxonomies and ontologies to help mature, extend, and evolve SE knowledge~\cite{usman2017taxonomies}. The Guide to the Software Engineering Body of Knowledge (SWEBOK) can be considered as a taxonomy that classifies software engineering discipline and its body of knowledge in a structured way~\cite{bourque2014guide}. Software engineering knowledge areas are defined in SWEBOK, and they can be used as a structured way of communication in the discipline. Other examples of taxonomies in software engineering are the work of Glass et~al.~\cite{glass2002research} and Blum~\cite{blum1994taxonomy}. Specialised taxonomies with narrower scopes are also popular in the field. These taxonomies are focused on specific sub-fields of software engineering such as Taxonomy of IoT Client Architecture~\cite{taivalsaari2018taxonomy}, Taxonomy of Requirement Change~\cite{saher2017requirement}, Taxonomy of Architecture Microservices~\cite{garriga2017towards}, Taxonomy of Global Software Engineering~\cite{vsmite2014empirically}, and Taxonomy of Variability Realisation Techniques~\cite{svahnberg2005taxonomy} to name a few.

This paper presents a taxonomy of assets in the inception, planning, development, evolution, and maintenance of a software-intensive product or service. The taxonomy is built using a hybrid method, i.e., the combination of top-down and bottom-up. The details of the taxonomy creation are presented in Section~\ref{sec:3_sub_taxonomy_creation}.

\subsection{Summary of the Gaps}\label{sec:sub_related_gap}

In this paper, we identify and categorise assets in software development and software engineering. We use the concept of assets and their intentional and unintentional degradation. Moreover, we aim to address the following real-world problems:

\begin{itemize}
    
    \item Bring awareness to practitioners and academicians. A precise and concise terminology of assets enables practitioners and academicians to consider new ways of dealing with asset degradation~\cite{zabardast2022assets}.
  
    \item The ripple effect that the degradation of an asset can impose on other assets is another aspect that necessitates the creation of taxonomy to understand relations between assets~\cite{kruchten2019managing,zabardast2022assets}. 
\end{itemize}

In this paper, we identify the software artefacts that adhere to this definition of assets and are common in the industry. We aim to address the following gap: identifying and distinguishing assets by considering every aspect of software development. For example, assets related to\textit{environment and infrastructure}, \textit{development Process}, \textit{ways of working}, and \textit{organisational aspects}~\cite{avgeriou2016managing,rios2018tertiary} that have not received enough attention. We aspire to identify assets and provide a synthesis of existing knowledge in the area of asset management.

\section{Research Overview}\label{sec:research_overview}

This section presents the research methodology of the paper. The process followed to build the taxonomy is divided into two main parts: a systematic literature review (SLR) and a field study~\cite{Stol2018} of industrial cases using focus group interviews and field notes as data collection methods. Combining an SLR and a field study helps us look at the phenomenon from different perspectives and create a complete picture of the phenomena of assets. In addition, this multi-faceted view provides insights that can strengthen the validity of the results.

In this work, we are answering the following research question:

\textbf{$RQ:$} \textit{What assets are managed by organisations during the inception, planning, development, evolution, and maintenance of software-intensive products or services?}

The rest of this section presents the SLR  (See Section~\ref{sec:3_sub_literature_review}), the field study~\cite{Stol2018} (See Section~\ref{sec:3_sub_industrial_ws}), and taxonomy creation (See Section~\ref{sec:3_sub_taxonomy_creation}).

\subsection{Systematic Literature Review: Planning and Execution}\label{sec:3_sub_literature_review}

This subsection describes the systematic literature review conducted in this work. 

All assets are subject to degradation. Asset degradation can be classified into three categories deliberate, unintentional, and entropy~\cite{zabardast2022assets}. Asset degradation can also be measured and monitored using different metrics, e.g., the amount of TD. TD is a familiar concept for practitioners and has received a growing interest in academia and the industry~\cite{rios2018tertiary}. TD has become a broad research area that has focused on different assets. The fact that TD research studies a particular asset might imply that that asset might be of value for software development organisations, or at least that might be perceived as such by TD researchers. Therefore, we do believe we can use the TD literature as a proxy for identifying and categorising different types of assets.

In order to study the state of art, we performed an SLR to capture the classifications and definitions of the various assets addressed by these previous research works (top-down method)~\cite{broughton2015essential}. In our literature review, the goal was to identify systematic literature reviews, systematic mapping studies, and tertiary studies. We reviewed secondary and tertiary research papers by performing an SLR using snowballing, following the guidelines by Wohlin~\cite{wohlin2014guidelines}. We selected snowballing as a search strategy as it allowed us to explore the area as well as its reported efficiency~\cite{badampudi2015experiences}.

The starting set of papers for snowballing was collected through a database search in Google Scholar in October 2021 using the following search string in: \textit{``technical debt'' AND (``systematic literature review'' OR ``systematic mapping study'' OR ``tertiary study'')}. We chose to use the search string only in Google scholar since it is not restricted to specific publishers~\cite{badampudi2015experiences}, and it can help avoid publisher bias~\cite{wohlin2014guidelines}. We selected, with the inclusion and exclusion described below, the articles that presented a classification for TD, i.e., articles that present different types of TD.

\begin{figure*}[ht]
\centering
  \includegraphics[width=1\textwidth]{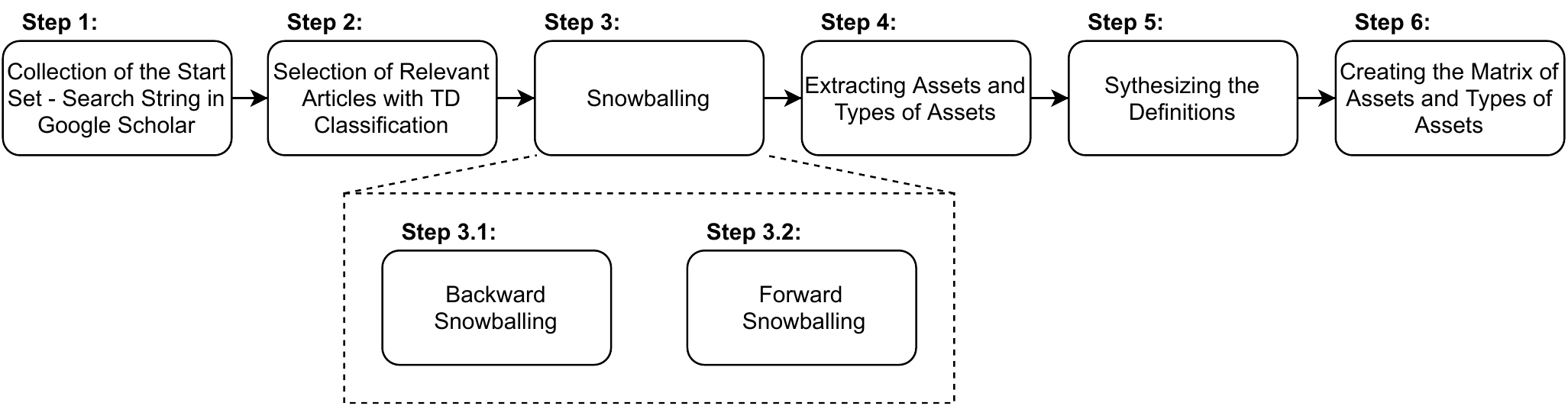}
\caption{The literature review execution process.}
\label{fig:01_lr_process}   
\end{figure*}

The SLR execution procedure to identify assets included the following steps as illustrated in Figure~\ref{fig:01_lr_process}. The results of this process are presented in Section~\ref{sec:4_sub_results_literature_review}.

\begin{itemize}
    \item \textbf{Step 1:} Collection of the start set of relevant articles (seed papers), including SLRs, SMSs, and tertiary studies on TD, by using a search string.
    \item \textbf{Step 2:} Evaluate the papers - start set in the first iteration - for inclusion/exclusion based on the criteria.
    \begin{itemize}
        \item \textbf{Inclusion Criteria:} 
        \begin{itemize}
        \item[$\circ$] The selected papers should report secondary or tertiary studies;
        \item[$\circ$] the papers should present any classification of TD and assets affected by TD;
        \item[$\circ$] the papers should be written in English; the papers' main aim is the literature review.
        \end{itemize}
        \item \textbf{Exclusion Criteria:} 
        \begin{itemize}
        \item[$\circ$] The papers that present informal literature reviews and are duplications of previous studies are not included.
        \end{itemize}
    \end{itemize}
    \item \textbf{Step 3:} The snowballing procedure for identifying additional secondary and tertiary studies on TD that satisfy our inclusion/exclusion criteria:
    \begin{itemize}
        \item \textbf{Step 3.1:} Backward snowballing by looking at the references of the selected papers. The backward snowballing was finished in one round.
        \item \textbf{Step 3.2:} Forward snowballing by looking at the papers that cite the selected papers. The forward snowballing was finished in one round.
    \end{itemize}
    \item \textbf{Step 4:} Extracting different \textit{types of assets} and \textit{assets} together with their respective definitions from the selected articles.
    \item \textbf{Step 5:} Synthesising the definitions of the \textit{types of assets} and \textit{assets} provided by the selected articles.
    \item \textbf{Step 6:} Creating the matrix of \textit{types of assets} and \textit{assets} based on TD classifications defined by the selected articles.
\end{itemize}

\subsection{Field Study (Focus Group Interviews): Planning and Execution}\label{sec:3_sub_industrial_ws}
To study the state of practice, we performed field studies, using focus groups and field notes as data collection in five companies to find evidence on how assets are defined and used. The five companies were selected using convenience sampling, as the companies are involved in an ongoing research project that focuses, among other topics, on addressing asset degradation challenges.

\begin{figure*}[htpb]
\centering
  \includegraphics[width=1\textwidth]{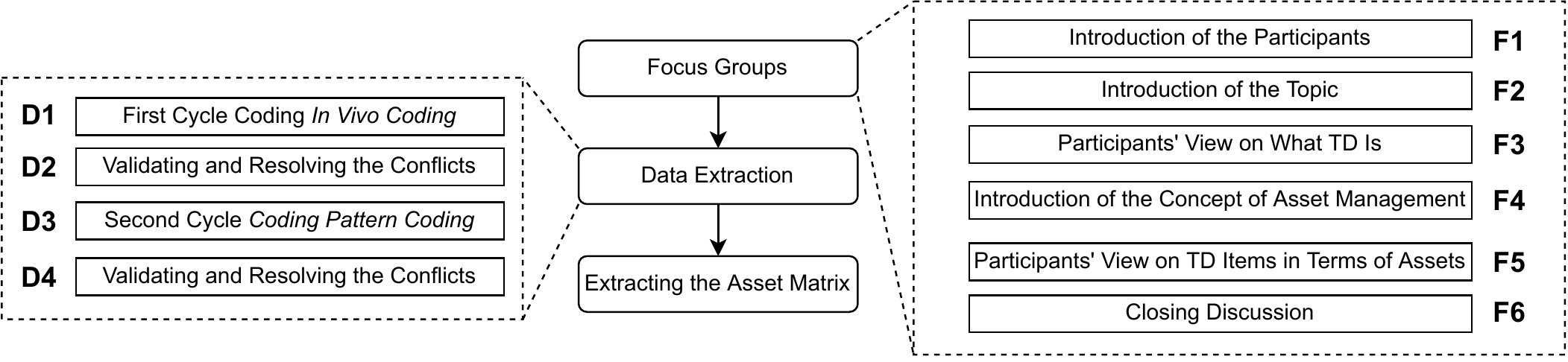}
\caption{The field study execution process.}
\label{fig:02_industrial_process}   
\end{figure*}

The focus groups’ process is presented in Figure~\ref{fig:02_industrial_process}. The reports from the focus groups were coded and used for the construction of the taxonomy. We used the bottom-up method~\cite{broughton2015essential} for updating the existing structure that we obtained from the literature review. The focus groups were planned as a half-working day (four hours). 

\subsubsection{Case Company Characterisation}

We have collected the data for this study by collaborating with five companies that work in the area of construction machinery, communication \& ICT, consultancy services, and financial services. The research partner companies are Ericsson (telecommunication \& ICT), Fortnox (Financial Services), Qtema (Consultancy Services), Time People Group (Consultancy Services), and Volvo CE (Construction Machinery). 

The partner companies are mature in their development practices and have well-established, successful products. They are interested in continuously improving their products and development processes, which turns into their willingness to participate in studies like this. All the collaborating companies work on developing software-intensive products or services and are involved in a large ongoing research partnership\footnote{See \url{www.rethought.se}.}. The details of the case companies are presented in Table~\ref{tab:tab1_case_inf}. Note that the order of the companies in Table~\ref{tab:tab1_case_inf} does not correspond with the order of focus groups (focus group IDs) in Table~\ref{tab:tab4_ws_asset_matrix}, which has been shuffled to preserve confidentiality.

\begin{table*}[htpb]
\caption{Case company details. The table is ordered alphabetically based on the name of the companies, and does not correspond to the order in Table \ref{tab:tab4_ws_asset_matrix}.}
\begin{tabular}{cccccc}
\toprule
\textbf{Company} & \textbf{Domain} & \textbf{\makecell{Investigated\\Site}} & \textbf{\makecell{Enterprise\\Size}} & \textbf{\makecell{Participants'\\Roles}} & \textbf{\makecell{Number of\\Participants}} \\
\midrule
Ericsson & \makecell{Telecommunication\\\& ICT} & \makecell{Karlskrona,\\Sweden} & Large & \makecell{Senior System Architect\\Corporate Senior\\Strategic Expert\\Operations \& Testing} & 3 \\
\midrule
Fortnox & Finance & \makecell{V\"axj\"{o},\\Sweden} & Large & \makecell{Head of Development\\Product Owner\\Development Manager\\System Architect\\Testing} & 14 \\
\midrule
\makecell{Qtema$\dagger$} & Consultancy & \makecell{Stockholm,\\Sweden} & SME & \makecell{Chairman of the Board\\Requirements Analyst\\Sales Manager\\Project Manager\\IT Administration\\Manager} & 6 \\
\midrule
\makecell{Time People Group$\dagger$} & Consultancy & \makecell{Stockholm,\\Sweden} & SME & \makecell{Data Consultant\\Project Manager\\Consultant\\Senior Agile Coach\\IT Project Manager\\Team Leader\\Chief Executive\\Officer (CEO)\\Consultant\\Test Leader} & 7 \\
\midrule
Volvo CE & \makecell{Construction\\Machinery} & \makecell{Gothenburg,\\Sweden} & Large & \makecell{Enterprise Architect\\Solution Architect\\Business Information\\Architect} & 5 \\
\bottomrule
\multicolumn{6}{l}{$\dagger$ Time People Group and Qtema participated in the same workshop.}
\end{tabular}

\label{tab:tab1_case_inf}
\end{table*}

\subsubsection{Focus Groups' Procedure}

The steps taken during the industrial focus groups are presented in this section. The focus groups include six steps (see Figure~\ref{fig:02_industrial_process}):
\begin{itemize}
    \item \textbf{\textit{F1.}} Focus group participants introducing themselves.
    \item \textbf{\textit{F2.}} One of the moderating researchers presenting the topic.
    \item \textbf{\textit{F3.}} Focus group participants discussing the topic, providing insight into their views/experiences with TD and document them in notes.
    \item \textbf{\textit{F4.}} Focus group participants discussing assets and asset management in detail after a second presentation of the concept.
    \item \textbf{\textit{F5.}} Focus group participants discussing what they wrote before as TD examples and rethinking them in terms of assets, asset degradation, and asset management.
    \item \textbf{\textit{F6.}} A closing discussion and focus groups.
\end{itemize}

Each Focus group starts with participants introducing themselves with background information about their work, including their current role in the organisation (step F1). One of the moderating researchers then presents the Focus group's agenda and covers the importance of the topic and the growing interest in value creation and waste reduction both in academia and in the industry (step F2).

After the initial introduction of the topic by the moderating researchers, the participants are divided into groups. They are asked to list and discuss the challenges with their ways of working (while considering varying aspects of TD), i.e., the problems they know or have encountered or experienced (step F3). After the time is up, the notes are read, discussed, and abstracted to a more general description and later put on a whiteboard. The connections between the items on the board are identified and marked down with a marker. 

After a second presentation, i.e., introducing the participants to the concepts related to asset management and asset degradation (step F4), the participants add new items to the previous notes on the board. Participants then refine the items from the board for the rest of the Focus group (step F5). The Focus group ends with a closing discussion on the topic and the items (step F6).

In the context of this research, we have moved the focus from the traditional TD metaphor to asset degradation. In this framework, we talk about asset degradation as the deviation of an asset from its representation. That way, we can focus, potentially, on any type of asset and its representation. This framework provides us with a broader, holistic view that allows us to study how an asset's degradation (e.g., requirements) might introduce degradation in other assets (e.g., code or test cases). 

The researchers' minutes that were written during each session were then aggregated and summarised in a report sent back to the participants, that were asked to provide us with feedback. The written notes from the participants and the final reports were used as raw data for creating the taxonomy using the data extraction method described in Section~\ref{sec:3_subsub_data_extraction}.  The raw data (i.e., participants' notes and the reports) were used for coding and later to extract \textit{types of assets} and explicit \textit{assets}. The details of the data extraction and taxonomy creation are described in Section~\ref{sec:3_sub_taxonomy_creation}.

Unfortunately, since this research is under non-disclosure agreements (NDA) with participating companies, we cannot disclose any further information about the companies, the participants, or the collected materials.

\subsubsection{Data Extraction}\label{sec:3_subsub_data_extraction}

To create the matrix of assets from industrial insights, we use the hybrid method of coding, as suggested by Salda\~{n}a~\cite{saldana2015coding}. The coding is divided into two main cycles: First Cycle Coding and Second Cycle Coding. 

\textbf{\textit{First Cycle Coding}} of the raw data happens in the initial stage of coding. The raw data, which can be a clause, a sentence, a compound sentence, or a paragraph, is labelled based on the semantic content and the context in which it was discussed during the Focus group. We have used the \textit{in vivo coding} method~\cite{saldana2015coding} to label the raw data in the first cycle. In vivo coding prioritises the participants' opinions~\cite{saldana2015coding}; therefore, it is suitable for labelling raw data in the first cycle coding in our study, where participants' opinions are used as input. It adheres to the ``verbatim principle, using terms and concepts drawn from the words of the participants themselves. By doing so, [the researchers] are more likely to capture the meanings inherent to people’s experiences.''~\cite[p.~140]{stringer2014} It is commonly used in empirical and practitioner research.~\cite{coghlan2014doing,fox2007doing,stringer2014}

The coding was done by two researchers independently (step D1, see Figure~\ref{fig:02_industrial_process}). The labels were then compared to validate the labels and to identify conflicting cases. A third researcher helped to resolve the conflicts by discussing the labels with the two initial researchers (step D2, see Figure~\ref{fig:02_industrial_process}).

\textbf{\textit{Second Cycle Coding}} is done primarily to categorise, theorise, conceptualise, or reorganise the coded data from the first cycle coding. We have used Pattern Coding~\cite{saldana2015coding} as the second cycle coding method. Pattern codes are explanatory or inferential codes that identify an emergent theme, configuration or explanation~\cite[p.~237]{saldana2015coding}. According to Miles et~al.~\cite[p.~86]{miles2014qualitative}, pattern coding is used in cases where: 
\begin{inparaenum} [(i)]
    \item the researchers aim to turn larger amounts of data into smaller analytical units.
    \item the researchers aim to identify themes from the data.
    \item the researchers aim to perform cross-case analysis on common themes from the data gathered by studying multiple cases.
\end{inparaenum}

Similar to the first cycle coding process, pattern coding was done by two researchers independently (step D3, see Figure~\ref{fig:02_industrial_process}). The results were compared to validate the classifications and to identify conflicting cases. A third researcher resolved the conflicting cases in a discussion session with the two researchers (step D4, see Figure~\ref{fig:02_industrial_process}). The results of the insights gathered from industrial focus groups are presented in Section~\ref{sec:4_sub_results_ws}.

\subsection{Taxonomy Creation}\label{sec:3_sub_taxonomy_creation}

\begin{figure}[ht]
\centering
  \includegraphics[width=0.47\textwidth]{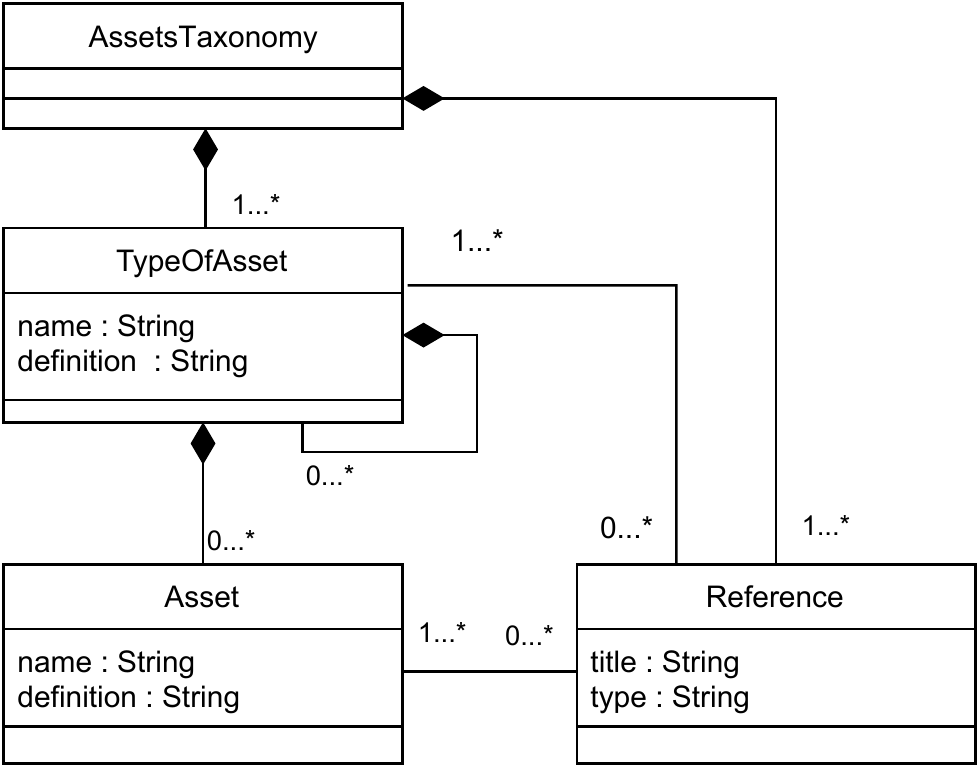}
\caption{Asset Management Metamodel.}
\label{fig3:metamodel}  
\end{figure}

To describe a precise syntax and the semantics of the different concepts used for the taxonomy creation, we created a metamodel (presented in Figure~\ref{fig3:metamodel}). The metamodel illustrates the structural relationships between the metaclasses, i.e., concepts, in the taxonomy. The metaclasses presented in the metamodel are:

\begin{itemize}
    \item The \textit{``AssetsTaxonomy''} metaclass is the container metaclass for the items in the model.
    \item The ``TypeOfAsset'' metaclass represents the hierarchical classification of assets. The items belonging to this metaclass can be further broken down into subclassifications representing various groups of assets. The types of assets are containers for the assets. Types of assets are identified from the state-of-the-art (i.e., existing academic literature), state-of-practice (i.e., the industrial insights gathered through the industrial focus groups), or the identified by researchers.
    \item The \textit{``Asset''} metaclass represents assets. Each asset belongs to one and only one type of asset, assuring orthogonality by design. Assets are identified from the state-of-the-art (i.e., existing academic literature), state-of-practice (i.e., the industrial insights gathered through the industrial focus groups), or identified by researchers.
    \item The \textit{``Reference''} metaclass represents the references from which each asset or type of asset has been identified. Reference can originate from academic literature (the literature review) or industrial insights (gathered from industrial focus groups). References can be mapped to individual \textit{assets/type of assets} or multiple \textit{assets/type of assets}.
\end{itemize}

The creation of the taxonomy included three steps. First, we summarised the relevant topics on TD types (top-down approach) with items that we extracted from the literature review. We created the matrix based on the literature's definitions, i.e., by synthesising the definitions to identify similarities, differences, and hierarchies of identified items. We have grouped the definitions provided by the literature based on their semantic meaning. 

In the second step, we utilised the extracted assets from industrial focus groups (Section~\ref{sec:3_subsub_data_extraction}) to create a second asset matrix (bottom-up approach). Like the previous step, we used the definitions of the assets and their types and the participants' statements from the focus groups.

By combining the two matrices in the last step, we created the asset management tree. To complete the tree, we added some nodes based on the researchers' expertise and extensive industry knowledge from decades of industrial research that we perceived were missing nodes and leaves. We mention such cases as \textit{Author Defined Assets} (ADA) when presenting the results.

The process of adding ADA started with researchers suggesting assets that should be included in the taxonomy. These suggested assets were brought up in internal workshops\footnote{Internal workshops refer to the workshops where the researchers discussed the taxonomy.} where all the researchers discussed, reflected, improved, and added or removed the suggested assets. \textit{User Stories}, as an example, were suggested by one of the researchers to be considered as assets during one of the internal workshops. The discussion was regarding
\begin{inparaenum}[(i)]
    \item whether or not \textit{User Stories} are assets,
    \item if they fit in the taxonomy according to the definition,
    \item where they belong in the tree,
    \item what they represent,
    \item and how they degrade.
\end{inparaenum}
After the discussions, the researchers decided that \textit{User Stories} belong to \textit{[AM1]} - \textit{[AM1.1] Functional-Requirements-Related Assets} in the taxonomy tree. The assets included in the taxonomy should adhere to the definition of an asset~\cite{zabardast2022assets}.

\subsection{Taxonomy Validation}\label{sec:3_sub_internal_validation}

This section describes the taxonomy validation procedure. According to Usman et~al.~\cite{usman2017taxonomies}, the validation process includes orthogonality demonstration, benchmarking, and utility demonstration. The taxonomy was created to be orthogonal by design (i.e., each element can only be a member of one group), as described in Section~\ref{sec:3_sub_taxonomy_creation}. Therefore, in this validation we focus on benchmarking and utility demonstration. 

We conducted three separate workshops (one for each company) to validate the taxonomy and its structure with the six participants (including Product Owner, Developer, Software Architect, Scrum Master, Test Quality Assurance, and Research Engineer) from three companies namely, Ericsson (2 participants), Fortnox (3 participants), and Volvo CE (1 participant) who were involved during the industrial workshops for the data collection. The procedure for the validation workshops is presented below.

\begin{enumerate}
    \item The taxonomy was sent to the participants to study before the validation workshops.
    \item During the validation workshops, the taxonomy was presented to the participants.
    \item After the presentation, the participants filled in a questionnaire that included four questions:
    \begin{enumerate}
        \item ``\textit{Select the assets from the taxonomy that you were not aware of.}''
        \item ``\textit{Select the assets that you think should not be in the taxonomy.}''
        \item ``\textit{Write down the assets that are missing from the taxonomy.}''
        \item ``\textit{Prioritise the top 5 assets based on your experience. (index starting from 1 as the most important asset). You can add missing assets from the previous question as well.}''
    \end{enumerate}
    \item At the end of each workshop, a discussion session was held where participants asked questions regarding the taxonomy and its content. The discussions were recorded.
\end{enumerate}

We decided to apply changes to the taxonomy after the validation workshops were concluded, in cases when the majority of the participants suggested a given change in the taxonomy. The results of the validation workshops are presented in Section~\ref{sec:validation_results}.

\section{Results}\label{sec:results}

This section presents the results of the systematic literature review and then the results from the field study. We will present the asset management taxonomy built by aggregating the results from both. Finally, we will present the results of the validation workshops.

\subsection{Results from the SLR}\label{sec:4_sub_results_literature_review}

The final list of selected papers included nine articles, presented in Table~\ref{tab:tab2_lr_papers}. To create the final assets' matrix (presented in Section~\ref{sec:4_AM_taxonomy}), we extracted the data from each article. For example, in paper \textit{P4}~\cite{li2015systematic}, we refer to Figure~8 on page ten of the article, where the authors summarise the ``TD classification tree'' and their respective definitions. The authors define Requirements TD as ``the distance between the optimal requirements specifications and the actual system implementation, under domain assumptions and constraints''~\cite[p. 9]{li2015systematic}. Requirements is also mentioned as a TD item in \textit{P3}, \textit{P5}, \textit{P6}, and \textit{P8}. Table~\ref{tab:tab3_lr_asset_matrix} presents the summary of our findings based on the types of TD. The codes of the papers are used as a reference throughout this paper. It is important to mention that Table~\ref{tab:tab3_lr_asset_matrix} does not represent an overview of the final assets but only the ones extracted from the SLR. However, this matrix helps us to categorise assets and types of assets according to existing classifications.

Looking at Table~\ref{tab:tab3_lr_asset_matrix} we can see that there are fewer categories in the earlier studies (\textit{P1} and \textit{P2}) compared to later studies (\textit{P3} to \textit{P9}). The more recent papers that follow these studies break down the bigger categories into more specific categories. For example, \textit{Architecture and Design} are put into one category in \textit{P1} and \textit{P2} and later, they are broken down into their own categories. It is important to mention that \textit{P7} has fewer categories since the study is on the specific topic of \textit{Architectural Technical Debt}.

\begin{table*}[htpb]
\caption{The articles gathered for the literature review during the snowballing process.}
\begin{tabular}{cp{10.25cm}ccc}
\toprule
\textbf{Code} & \textbf{Title} & \textbf{\makecell{Seed\\Paper}} & \textbf{\makecell{Backward\\Snowballing}} & \textbf{\makecell{Forward\\Snowballing}} \\
\midrule
P1 & A Consolidated Understanding of Technical Debt~\cite{tom2012consolidated} &&x& \\
P2 & An Exploration of Technical Debt~\cite{tom2013exploration} &&x& \\
P3 & Towards an Ontology of Terms on Technical Debt~\cite{alves2014towards} &x&& \\
P4 & A Systematic Mapping Study on Technical Debt and Its Management~\cite{li2015systematic} &x&& \\
P5 & Identification and Management of Technical Debt: A Systematic Mapping Study~\cite{alves2016identification} &x&& \\
P6 & Managing Architectural Technical Debt: A Unified Model and Systematic Literature Review~\cite{besker2018managing} &x&& \\
P7 & A tertiary study on technical debt: Types, management strategies, research trends, and base information for practitioners~\cite{rios2018tertiary} &x&& \\
P8 & A systematic literature review on Technical Debt prioritization: Strategies, processes, factors, and tools~\cite{lenarduzzi2019technical} &&&x \\
P9 & Investigate, identify and estimate the technical debt: a systematic mapping study~\cite{benidris2020investigate} &&&x \\

\bottomrule
\end{tabular}
\label{tab:tab2_lr_papers}
\end{table*}
\begin{table*}[htpb]
\caption{Asset matrix from technical debt literature.}
\scriptsize
\begin{adjustbox}{angle=90}
\begin{tabular}{c|c|c|c|c|c|c|c|c|cc}
\textbf{\makecell{P1\\2012}} & \textbf{\makecell{P2\\2013}} & \textbf{\makecell{P3\\2014}} & \textbf{\makecell{P4\\2015}} & \textbf{\makecell{P5\\2016}} & \textbf{\makecell{P6\\2018}} & \textbf{\makecell{P7\\2018}} & \textbf{\makecell{P8\\2019}} & \textbf{\makecell{P9\\2020}} & \textbf{\makecell{Emerging\\Category(ies)}} & \textbf{\makecell{Phase,\\during\\which the\\artefact is\\produced}} \\
\midrule
\makecell{\tikz\draw[blue,fill=blue] (0,0) circle (1ex); Features} & & \makecell{\tikz\draw[green,fill=green] (0,0) circle (1ex);\\Requirements} & \makecell{\tikz\draw[green,fill=green] (0,0) circle (1ex);\\Requirements} & \makecell{\tikz\draw[green,fill=green] (0,0) circle (1ex);\\Requirements} & \makecell{\tikz\draw[green,fill=green] (0,0) circle (1ex);\\Requirements} & & \makecell{\tikz\draw[green,fill=green] (0,0) circle (1ex);\\Requirements} & \makecell{\tikz\draw[green,fill=green] (0,0) circle (1ex);\\Requirements} & \textit{\makecell{\tikz\draw[green,fill=green] (0,0) circle (1ex); Product\\Requirements}} & \multirow{2}{*}{Requirements} \\
 & & & & \makecell{\tikz\draw[green,fill=green] (0,0) circle (1ex); Usability} & \makecell{\tikz\draw[green,fill=green] (0,0) circle (1ex); Usability} & & & \makecell{\tikz\draw[green,fill=green] (0,0) circle (1ex); Usability} & \textit{\makecell{\tikz\draw[green,fill=green] (0,0) circle (1ex); Quality\\Requirements}} & \\
 \midrule
\makecell{\tikz\draw[green,fill=green] (0,0) circle (1ex); Design\textbackslash\\Architecture} & \makecell{\tikz\draw[green,fill=green] (0,0) circle (1ex); Design and \\Architecture} & \makecell{\tikz\draw[blue,fill=blue] (0,0) circle (1ex); Architecture\\Design} & \makecell{\tikz\draw[blue,fill=blue] (0,0) circle (1ex); Architecture\\Design} & \makecell{\tikz\draw[blue,fill=blue] (0,0) circle (1ex); Architecture\\Design} & \makecell{\tikz\draw[blue,fill=blue] (0,0) circle (1ex); Architecture\\Design} & \makecell{\tikz\draw[blue,fill=blue] (0,0) circle (1ex);\\Architecture} & \makecell{\tikz\draw[blue,fill=blue] (0,0) circle (1ex); Architecture\\Design} & \makecell{\tikz\draw[blue,fill=blue] (0,0) circle (1ex); Architecture\\Design} & \textit{\makecell{\tikz\draw[blue,fill=blue] (0,0) circle (1ex); Architecture\\Design\\Decisions\\Documentation}} & \multirow{4}{*}{Design} \\
\makecell{\tikz\draw[green,fill=green] (0,0) circle (1ex);\\Documentation} & \makecell{\tikz\draw[green,fill=green] (0,0) circle (1ex);\\Documentation} & \makecell{\tikz\draw[green,fill=green] (0,0) circle (1ex);\\Documentation} & \makecell{\tikz\draw[green,fill=green] (0,0) circle (1ex);\\Documentation} & \makecell{\tikz\draw[green,fill=green] (0,0) circle (1ex);\\Documentation} & \makecell{\tikz\draw[green,fill=green] (0,0) circle (1ex);\\Documentation} & & \makecell{\tikz\draw[green,fill=green] (0,0) circle (1ex);\\Documentation} & \makecell{\tikz\draw[green,fill=green] (0,0) circle (1ex);\\Documentation} & \textit{\makecell{\tikz\draw[blue,fill=blue] (0,0) circle (1ex); Product\\Documentation}} &\\
 & & \makecell{\tikz\draw[blue,fill=blue] (0,0) circle (1ex); Design\\Specifications} &&&&& \makecell{\tikz\draw[blue,fill=blue] (0,0) circle (1ex);\\Architectural\\Documentation} && \textit{\makecell{\tikz\draw[blue,fill=blue] (0,0) circle (1ex);\\Architectural\\Documentation} } &\\
 \midrule
\makecell{\tikz\draw[blue,fill=blue] (0,0) circle (1ex); Code} & \makecell{\tikz\draw[blue,fill=blue] (0,0) circle (1ex); Code} & \makecell{\tikz\draw[blue,fill=blue] (0,0) circle (1ex); Code} & \makecell{\tikz\draw[blue,fill=blue] (0,0) circle (1ex); Code} & \makecell{\tikz\draw[blue,fill=blue] (0,0) circle (1ex); Code} & \makecell{\tikz\draw[blue,fill=blue] (0,0) circle (1ex); Code} & \makecell{\tikz\draw[blue,fill=blue] (0,0) circle (1ex); Code} & \makecell{\tikz\draw[blue,fill=blue] (0,0) circle (1ex); Code} & \makecell{\tikz\draw[blue,fill=blue] (0,0) circle (1ex); Code} & \makecell{\tikz\draw[blue,fill=blue] (0,0) circle (1ex);\\\textit{Source Code}} & \multirow{4}{*}{Development} \\
 & & \makecell{\tikz\draw[blue,fill=blue] (0,0) circle (1ex); Build} & \makecell{\tikz\draw[green,fill=green] (0,0) circle (1ex); Build} & \makecell{\tikz\draw[green,fill=green] (0,0) circle (1ex); Build} & \makecell{\tikz\draw[green,fill=green] (0,0) circle (1ex); Build} & & \makecell{\tikz\draw[green,fill=green] (0,0) circle (1ex); Build} & \makecell{\tikz\draw[green,fill=green] (0,0) circle (1ex); Build} & \textit{\makecell{\tikz\draw[green,fill=green] (0,0) circle (1ex); Build\\Documentation}} &\\
 & & \makecell{\tikz\draw[blue,fill=blue] (0,0) circle (1ex); Service} & & \makecell{\tikz\draw[blue,fill=blue] (0,0) circle (1ex); Service} & \makecell{\tikz\draw[blue,fill=blue] (0,0) circle (1ex); Service} & & & \makecell{\tikz\draw[blue,fill=blue] (0,0) circle (1ex); Service} & \textit{\makecell{\tikz\draw[blue,fill=blue] (0,0) circle (1ex); Web\\Services}} &\\
 & & & \makecell{\tikz\draw[green,fill=green] (0,0) circle (1ex);\\Versioning} & \makecell{\tikz\draw[green,fill=green] (0,0) circle (1ex);\\Versioning} & \makecell{\tikz\draw[green,fill=green] (0,0) circle (1ex);\\Versioning} & & \makecell{\tikz\draw[green,fill=green] (0,0) circle (1ex);\\Versioning} & \makecell{\tikz\draw[green,fill=green] (0,0) circle (1ex);\\Versioning} & \makecell{\tikz\draw[green,fill=green] (0,0) circle (1ex);\\\textit{Versioning}} &\\
 \midrule
\makecell{\tikz\draw[green,fill=green] (0,0) circle (1ex); Testing} & \makecell{\tikz\draw[green,fill=green] (0,0) circle (1ex); Testing} & \makecell{\tikz\draw[green,fill=green] (0,0) circle (1ex); Test} & \makecell{\tikz\draw[green,fill=green] (0,0) circle (1ex); Test} & \makecell{\tikz\draw[green,fill=green] (0,0) circle (1ex); Test} & \makecell{\tikz\draw[green,fill=green] (0,0) circle (1ex); Test} & & \makecell{\tikz\draw[green,fill=green] (0,0) circle (1ex); Test} & \makecell{\tikz\draw[green,fill=green] (0,0) circle (1ex); Test} & \textit{\makecell{\tikz\draw[green,fill=green] (0,0) circle (1ex); Functional\\Test}} & \multirow{4}{*}{\makecell{Verification\\and\\Validation}}\\
\makecell{\tikz\draw[black,fill=white] (0,0) circle (1ex); Defects} & & \makecell{\tikz\draw[black,fill=white] (0,0) circle (1ex); Defects} & \makecell{\tikz\draw[black,fill=white] (0,0) circle (1ex); Defects} & \makecell{\tikz\draw[black,fill=white] (0,0) circle (1ex); Defects} & \makecell{\tikz\draw[black,fill=white] (0,0) circle (1ex); Defects} & & \makecell{\tikz\draw[black,fill=white] (0,0) circle (1ex); Defects}& \makecell{\tikz\draw[black,fill=white] (0,0) circle (1ex); Defects}&&\\
& & \makecell{\tikz\draw[green,fill=green] (0,0) circle (1ex); Test\\Automation} & & \makecell{\tikz\draw[green,fill=green] (0,0) circle (1ex); Test\\Automation} & \makecell{\tikz\draw[green,fill=green] (0,0) circle (1ex); Test\\Automation} & & & \makecell{\tikz\draw[green,fill=green] (0,0) circle (1ex); Test\\Automation} & \textit{\makecell{\tikz\draw[green,fill=green] (0,0) circle (1ex); Test\\Automation}}&\\
&&&&\makecell{\tikz\draw[green,fill=green] (0,0) circle (1ex); Test Case\\Documentation}&& &&&\textit{\makecell{\tikz\draw[green,fill=green] (0,0) circle (1ex); Test\\Documentation}}&\\
\midrule
&\makecell{\tikz\draw[green,fill=green] (0,0) circle (1ex);\\Environment}&&&&& &&& \textit{\makecell{\tikz\draw[green,fill=green] (0,0) circle (1ex);\\Environment \&\\Infrastructure}} & \multirow{3}{*}{\makecell{Operations}} \\
\makecell{\tikz\draw[green,fill=green] (0,0) circle (1ex);\\Infrastructure} & \makecell{\tikz\draw[green,fill=green] (0,0) circle (1ex); Hardware\\Infrastructure} & \makecell{\tikz\draw[green,fill=green] (0,0) circle (1ex);\\Infrastructure} & \makecell{\tikz\draw[green,fill=green] (0,0) circle (1ex);\\Infrastructure} & \makecell{\tikz\draw[green,fill=green] (0,0) circle (1ex);\\Infrastructure}& \makecell{\tikz\draw[green,fill=green] (0,0) circle (1ex);\\Infrastructure}& &\makecell{\tikz\draw[green,fill=green] (0,0) circle (1ex);\\Infrastructure}& \makecell{\tikz\draw[green,fill=green] (0,0) circle (1ex);\\Infrastructure}& \textit{\makecell{\tikz\draw[green,fill=green] (0,0) circle (1ex);\\Environment \&\\Infrastructure}}&\\
&\makecell{\tikz\draw[green,fill=green] (0,0) circle (1ex); Operational\\Processes}&&&&& &&&\textit{\makecell{\tikz\draw[green,fill=green] (0,0) circle (1ex); Operations}}&\\
\midrule
&&\makecell{\tikz\draw[green,fill=green] (0,0) circle (1ex); Process}&&\makecell{\tikz\draw[green,fill=green] (0,0) circle (1ex); Process}&\makecell{\tikz\draw[green,fill=green] (0,0) circle (1ex); Process}& &&\makecell{\tikz\draw[green,fill=green] (0,0) circle (1ex); Process}&\textit{\makecell{\tikz\draw[green,fill=green] (0,0) circle (1ex); Process\\Managmenet}}& \multirow{2}{*}{\makecell{Management}}\\
&&&&\makecell{\tikz\draw[blue,fill=blue] (0,0) circle (1ex);\\Documentation\\Specifications}&& &&&\textit{\makecell{\tikz\draw[blue,fill=blue] (0,0) circle (1ex);\\Documentation\\Internal Rules \&\\Specifications}}&\\
\bottomrule
\multicolumn{11}{c}{}\\
\multicolumn{2}{c}{Color Guide:}&\multicolumn{2}{c}{\tikz\draw[blue,fill=blue] (0,0) circle (1ex); Assets}&\multicolumn{2}{c}{\tikz\draw[green,fill=green] (0,0) circle (1ex); Types of Assets}&\multicolumn{2}{c}{\tikz\draw[black,fill=white] (0,0) circle (1ex); Temporary Artefacts}&\multicolumn{3}{c}{}\\
\end{tabular}
\end{adjustbox}
\label{tab:tab3_lr_asset_matrix}
\end{table*}

\subsection{Results from the Field Study}\label{sec:4_sub_results_ws}

There were a total of four focus groups, each held with the participation of employees from different companies. The focus groups' procedure stayed the same, while the closing discussion of each focus group was on the topic of interest for that focus group's participants (the stakeholders). The topics included, but were not limited to, \textit{Lack of Knowledge/ Competence}, \textit{Architecture Lifecycle}, \textit{Business Models for Products}, and \textit{Backlog Update Issues/Backlog Size}.

Two researchers used the in vivo coding method to label $386$ statements during the first cycle coding. After matching and validating the labels, $14$ cases of conflicting labels were identified. The conflicting labels were resolved during a discussion session with a third researcher. The researchers agreed on the new labels for the conflicting cases during that discussion.

The focus groups are presented in chronological order in Table~\ref{tab:tab4_ws_asset_matrix}, i.e., \textit{WS1} was the first focus group. Examining Table~\ref{tab:tab4_ws_asset_matrix}, we observe that assets are mentioned more often than types of assets in the industrial focus groups, whereas types of assets are more frequent in the literature review (see Table~\ref{tab:tab3_lr_asset_matrix}). Finally, assets that are related to \textit{Operations}, \textit{Management}, and \textit{Organisational Management} were highlighted more in the industrial focus groups than the literature review.

\begin{table*}[htpb]
\caption{Asset matrix from industrial input.}
\scriptsize
\begin{tabular}{c|c|c|c|cc}
\textbf{WS1} & \textbf{WS2} & \textbf{WS3} & \textbf{WS4} & \textbf{\makecell{Emerging\\Category(ies)}} & \textbf{\makecell{Phase, during\\which the\\artefact is produced}} \\
\midrule
\makecell{\tikz\draw[green,fill=green] (0,0) circle (1ex); Contradictory\\ Requirements} &\makecell{\tikz\draw[green,fill=green] (0,0) circle (1ex); Requirements}&\makecell{\tikz\draw[green,fill=green] (0,0) circle (1ex); Requirements}&\makecell{\tikz\draw[green,fill=green] (0,0) circle (1ex); Requirements}&\textit{\makecell{\tikz\draw[green,fill=green] (0,0) circle (1ex); Product Requirements}}&Requirements\\
\midrule
&&\makecell{\tikz\draw[blue,fill=blue] (0,0) circle (1ex); Architectural\\Models} &&\tikz\draw[blue,fill=blue] (0,0) circle (1ex); Documentation&\\
&\tikz\draw[blue,fill=blue] (0,0) circle (1ex); Documentation &&\tikz\draw[blue,fill=blue] (0,0) circle (1ex); Documentation &\makecell{\tikz\draw[blue,fill=blue] (0,0) circle (1ex); Product\\Documentation}&\multirow{4}{*}{Design}\\
&\makecell{\tikz\draw[blue,fill=blue] (0,0) circle (1ex); Architectural\\Documents}&&\makecell{\tikz\draw[blue,fill=blue] (0,0) circle (1ex); Architectural\\Documentation}&\textit{\makecell{\tikz\draw[blue,fill=blue] (0,0) circle (1ex); Architectural\\Documentation}}&\\
&\tikz\draw[blue,fill=blue] (0,0) circle (1ex); Architecture&\tikz\draw[blue,fill=blue] (0,0) circle (1ex); Software Structure&\tikz\draw[blue,fill=blue] (0,0) circle (1ex); Architecture& \makecell{\tikz\draw[blue,fill=blue] (0,0) circle (1ex); Architectural\\(Source Code)}&\\
\midrule
\makecell{\tikz\draw[blue,fill=blue] (0,0) circle (1ex); Dangerous\\Code}&\tikz\draw[blue,fill=blue] (0,0) circle (1ex); Code&\tikz\draw[blue,fill=blue] (0,0) circle (1ex); Code&\tikz\draw[blue,fill=blue] (0,0) circle (1ex); Code&\tikz\draw[blue,fill=blue] (0,0) circle (1ex); Source Code& \multirow{3}{*}{Development}\\
&\tikz\draw[blue,fill=blue] (0,0) circle (1ex); APIs&\tikz\draw[blue,fill=blue] (0,0) circle (1ex); API Versions&&\tikz\draw[blue,fill=blue] (0,0) circle (1ex); \textit{APIs}&\\
&\tikz\draw[blue,fill=blue] (0,0) circle (1ex); Libraries&\makecell{\tikz\draw[blue,fill=blue] (0,0) circle (1ex); Third\\Party Products}&&\textit{\makecell{\tikz\draw[blue,fill=blue] (0,0) circle (1ex); Libraries\textbackslash\\External Libraries}}&\\
\midrule
&\tikz\draw[blue,fill=blue] (0,0) circle (1ex); Test Cases&\tikz\draw[blue,fill=blue] (0,0) circle (1ex); Tests&\tikz\draw[blue,fill=blue] (0,0) circle (1ex); Test Cases&\textit{\tikz\draw[blue,fill=blue] (0,0) circle (1ex); Test Cases}&\multirow{3}{*}{\makecell{Verification\\and\\Validation}}\\
&\tikz\draw[blue,fill=blue] (0,0) circle (1ex); Automated Tests&&&\textit{\makecell{\tikz\draw[blue,fill=blue] (0,0) circle (1ex); Test\\Automation Scripts}}&\\
&\tikz\draw[black,fill=white] (0,0) circle (1ex); Bug Reports&&&&\\
\midrule
&&&\tikz\draw[blue,fill=blue] (0,0) circle (1ex); Application Data&\textit{\tikz\draw[blue,fill=blue] (0,0) circle (1ex); Application Data}&\multirow{2}{*}{Operations}\\
&\tikz\draw[blue,fill=blue] (0,0) circle (1ex); Kubernets&\makecell{\tikz\draw[blue,fill=blue] (0,0) circle (1ex); Containers\\\textbackslash Kubernets}&&\textit{\tikz\draw[blue,fill=blue] (0,0) circle (1ex); Tools}&\\
\midrule
&\tikz\draw[blue,fill=blue] (0,0) circle (1ex); Ways of Working&\tikz\draw[blue,fill=blue] (0,0) circle (1ex); Ways of Working&\makecell{\tikz\draw[blue,fill=blue] (0,0) circle (1ex); Documentation\\about\\Ways of Working}&\textit{\makecell{\tikz\draw[blue,fill=blue] (0,0) circle (1ex); Documentation\\about\\Ways of Working}}&\multirow{6}{*}{Management}\\
&\tikz\draw[blue,fill=blue] (0,0) circle (1ex); Coding Standards&\tikz\draw[blue,fill=blue] (0,0) circle (1ex); Coding Standards&&\textit{\tikz\draw[blue,fill=blue] (0,0) circle (1ex); Coding Standards}&\\
&&\tikz\draw[blue,fill=blue] (0,0) circle (1ex); Architectural Rules&&\textit{\makecell{\tikz\draw[blue,fill=blue] (0,0) circle (1ex); Architectural\\Internal Standards}}&\\
&&&\makecell{\tikz\draw[blue,fill=blue] (0,0) circle (1ex); Documentation\\Standards}&\textit{\makecell{\tikz\draw[blue,fill=blue] (0,0) circle (1ex); Documentation\\Internal Rules\\\textbackslash Standards}}&\\
&\tikz\draw[blue,fill=blue] (0,0) circle (1ex); Product Roadmap&\tikz\draw[green,fill=green] (0,0) circle (1ex); Product Management&&\textit{\tikz\draw[green,fill=green] (0,0) circle (1ex); Product Management}&\\
&\tikz\draw[blue,fill=blue] (0,0) circle (1ex); Backlog&&&\textit{\tikz\draw[blue,fill=blue] (0,0) circle (1ex); Product Backlog}&\\
\midrule
&\makecell{\tikz\draw[blue,fill=blue] (0,0) circle (1ex); Organisation's\\Roadmap}&&\tikz\draw[blue,fill=blue] (0,0) circle (1ex); Holistic Strategy&\textit{\makecell{\tikz\draw[blue,fill=blue] (0,0) circle (1ex); Organisation's\\Strategy}}&\multirow{3}{*}{\makecell{Organisational\\Management}}\\
&&&\makecell{\tikz\draw[blue,fill=blue] (0,0) circle (1ex); Organisation's\\Structure}&\textit{\makecell{\tikz\draw[blue,fill=blue] (0,0) circle (1ex); Organisation's\\Structure}}&\\
&&\tikz\draw[blue,fill=blue] (0,0) circle (1ex); Business Models&&\textit{\tikz\draw[blue,fill=blue] (0,0) circle (1ex); Business Models}&\\
\bottomrule
\multicolumn{6}{c}{}\\
\multicolumn{2}{c}{Color Guide:}&\multicolumn{1}{c}{\makecell{\tikz\draw[blue,fill=blue] (0,0) circle (1ex); Assets}}&\multicolumn{1}{c}{\makecell{\tikz\draw[green,fill=green] (0,0) circle (1ex); Types of Assets}}&\multicolumn{1}{c}{\makecell{\tikz\draw[black,fill=white] (0,0) circle (1ex); Temporary Artefacts}}&\multicolumn{1}{c}{}\\
\end{tabular}
\label{tab:tab4_ws_asset_matrix}
\end{table*}

\subsection{The Asset Management Taxonomy}\label{sec:4_AM_taxonomy}

Using the key concepts extracted from the labelled data (presented in Table~\ref{tab:tab3_lr_asset_matrix} and Table~\ref{tab:tab4_ws_asset_matrix}), we build the taxonomy of assets. The taxonomy contains the assets identified both through the literature review and through the industrial focus groups. The assets included in the taxonomy are presented in a tree (graph). The nodes represent the assets (the leaf nodes) and the types of assets (non-leaf nodes).  

The tree presented in Figure~\ref{fig:04_AM_just_types} contains only the types of assets (The full tree is presented in Appendix~\ref{appendix}). Note that the nodes in the tree in Figure~\ref{fig:04_AM_just_types} are mapped to represent their source. For example, a node can be assigned with \textit{[P1]} as a reference for an article in the literature review. Similarly, \textit{[WS1]} as a reference for an asset coming from industrial focus groups\footnote{The IDs on the focus groups on Table~\ref{tab:tab4_ws_asset_matrix} have been obfuscated to preserve anonymity and have no relationship with the order of companies shown in Table~\ref{tab:tab1_case_inf}.}. And finally, Author Defined Assets (\textit{[ADA]}) are assets included in the taxonomy by the researchers.

\begin{figure*}[ht]
\centering
  \includegraphics[width=1\textwidth]{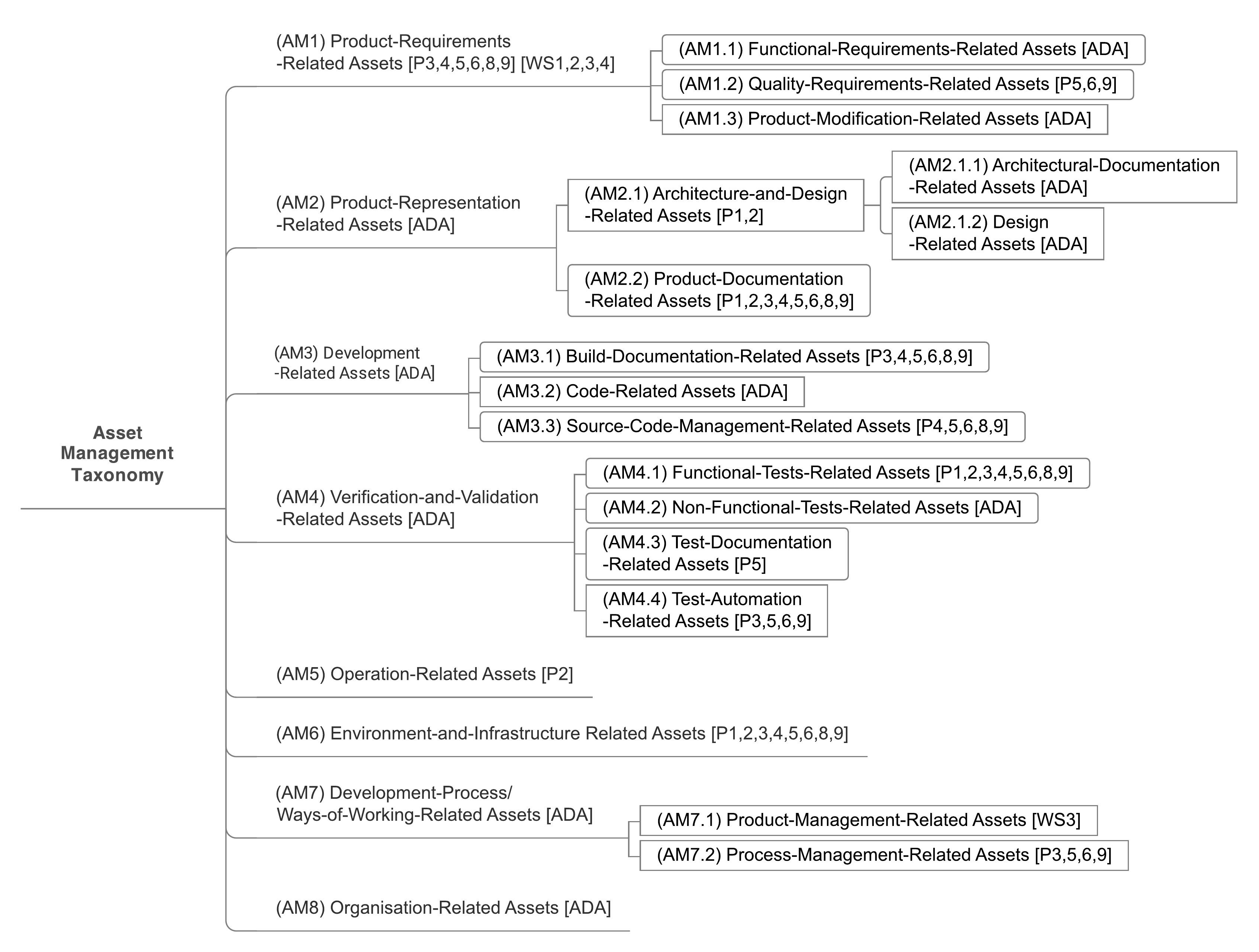}
\caption{The Asset Management Taxonomy. The tree contains only the types of assets. The full tree is presented in Appendix~\ref{appendix}.}
\label{fig:04_AM_just_types}   
\end{figure*}

The process described in Section~\ref{sec:3_sub_taxonomy_creation}, combining the asset matrices from the literature review, the input from the industrial focus groups, and completing the tree with \textit{Author Defined Assets} (ADA) resulted in the taxonomy containing $24$ types of assets and a total of $57$ assets. 

The eight main types of assets included in the taxonomy are:

\begin{itemize}

\item \textbf{Product-Requirements-Related Assets (AM1)} refer to assets (and types of assets) concerned with software requirements, including the elicitation, analysis, specification, validation, and management of requirements during the life cycle of the software product.

\item \textbf{Product-Representation-Related Assets (AM2)} refer to the assets (and types of assets) concerned with system and architectural design and any documentation related to these assets.

\item \textbf{Development-Related Assets (AM3)} refer to the assets (and types of assets) concerned with the development of the software product, including the code, build, and versioning.

\item \textbf{Verification-and-Validation-Related Assets (AM4)} refer to assets (and types of assets) concerned with software testing and quality assurance and the output provided by such assets that help the stakeholders investigate the quality of the software product.

\item \textbf{Operations-Related Assets (AM5)} refer to assets (and types of assets) concerned with the data produced or collected from operational activities, e.g., any data collected during the use of the product or service.

\item \textbf{Environment/Infrastructure-Related Assets (AM6)} refers to assets (and types of assets) concerned with the development environment, the infrastructure, and the tools (including support applications) that facilitate the development or deployment process. 

\item \textbf{Development-Process/Ways-of-Working-Related Assets (AM7)} refer to assets (and types of assets) concerned with product and process management and all the interrelated processes and procedures during the development process.

\item \textbf{Organisation-Related Assets (AM8)} refer to assets (and types of assets) concerned with organisations, such as team constellation, team collaborations, and organisational governance. 
 
\end{itemize}
In the remainder of the section, we present the eight major types of assets labelled \textit{AM1}-\textit{AM8}. We include the definitions of each type of asset together with their corresponding assets.

\subsubsection{Product-Requirements-Related Assets (AM1)}\label{sec:4_subsub_AM1}

\begin{figure*}[ht]
\centering
  \includegraphics[width=1\textwidth]{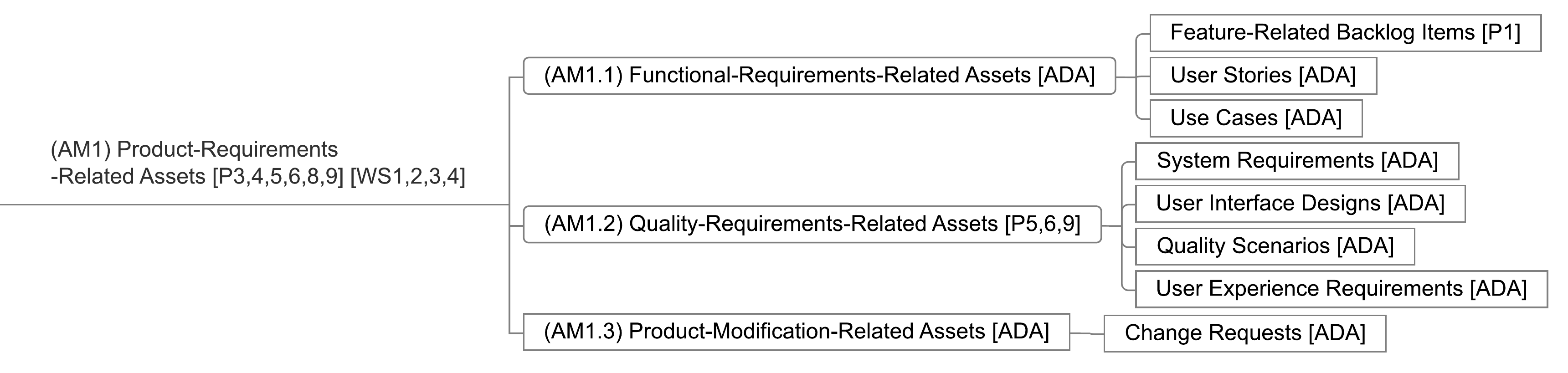}
\caption{Product-Requirements-Related Assets Sub-tree.}
\label{fig:05_AM1}   
\end{figure*}

Product-Requirements-Related Assets include the following three types of assets (see Figure~\ref{fig:05_AM1} with the taxonomy's sub-tree and Table~\ref{tab:tab5_am1} with definitions of each asset):

\begin{itemize}
    \item \textit{Functional-Requirement-Related Assets (AM1.1)} refer to the assets related to the functions that the software shall provide and that can be tested~\cite{bourque2014guide}. We have identified the following assets belonging to this type: \textit{Feature-Related Backlog Items}, \textit{User Stories}, and \textit{Use Cases}.
    \item \textit{Quality-Requirement-Related Assets (AM1.2)} refer to the assets related to non-functional requirements that act to constrain the solution~\cite{bourque2014guide}. We have identified the following assets belonging to this type: \textit{System Requirements}, \textit{User Interface Designs}, \textit{Quality Scenarios} (i.e., the -ilities), and \textit{User Experience Requirements}.
    \item \textit{Product-Modification-Related Assets (AM1.3)} refer to assets that mandate a change of the system and, but not necessarily, the requirements. \textit{Change Requests} is an asset we identified belonging to this type.
\end{itemize}

\begin{table*}[htpb]
\caption{Product-Requirements-Related Assets' Definitions.}
\begin{tabular}{p{2.5cm}p{1.2cm}p{12cm}}
\toprule
\textbf{Asset} & \textbf{AM type} & \textbf{Definition} \\
\midrule
Feature-Related Backlog Items & AM1.1 & Feature-Related Backlog Items are the results of refining and breaking down the user stories to create executable tasks~\cite{muter2019refinement}. \\
User Stories & AM1.1 & User Stories are, according to the agile development paradigm, a way to specify the features of the software that is being developed~\cite{muter2019refinement}. \\
Use Cases & AM1.1 & Use Cases are lists of actions or events that describe how a user will achieve a goal in a system~\cite{kruchten2004rational}.\\
System Requirements & AM1.2 & ``System Requirements are the requirements for the system as a whole. System Requirements [...] encompass user requirements, requirements of other stakeholders (such as regulatory authorities), and requirements without an identifiable human source.''~\cite{bourque2014guide}. \\
User Interface Designs & AM1.2 & ``User Interface Design is an essential part of the software design process. User interface design should ensure that interaction between the human and the machine provides for effective operation and control of the machine. For software to achieve its full potential, the user interface should be designed to match the skills, experience, and expectations of its anticipated users.''~\cite{bourque2014guide}. \\
Quality Scenarios (The -ilities) & AM1.2 & ``A quality attribute (QA) is a measurable or testable property of a system that is used to indicate how well the system satisfies the needs of its stakeholders.''~\cite{bass2003software} A quality scenario is a way of stating a requirement in an unambiguous and testable manner~\cite{bass2003software}. \\
User Experience Requirements & AM1.2 & User Experience Requirements ``are considered key quality determinants of any product, system or service intended for human use, which in turn can be considered as product, system or service success or failure indicators and improve user loyalty.''~\cite{law2010modelling, kujala2013emotions}. \\
Change Requests & AM1.3 & Change Requests are the modifications to the software product that are not coming from the requirements analysis of the product. \\
\bottomrule
\end{tabular}
\label{tab:tab5_am1}
\end{table*}

\subsubsection{Product-Representation-Related Assets (AM2)}\label{sec:4_subsub_AM2}

\begin{figure*}[ht]
\centering
  \includegraphics[width=1\textwidth]{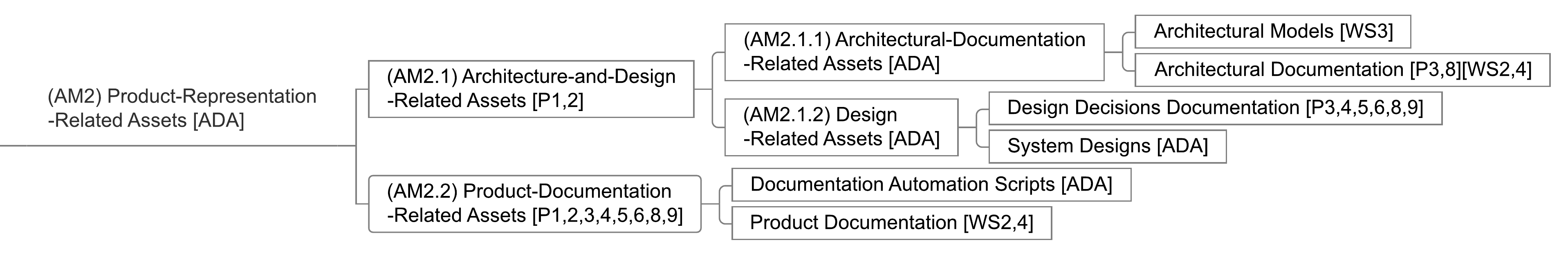}
\caption{Product-Representation-Related Assets Sub-tree.}
\label{fig:06_AM2}   
\end{figure*}

Product-Representation-Related Assets include the following two types of assets (see Figure~\ref{fig:06_AM2} with the sub-tree and Table~\ref{tab:tab6_am2} with definitions of each asset):

\begin{itemize}
    \item \textit{Architecture-and-Design-Related Assets (AM2.1)} refer to the assets that are used to design, communicate, represent, maintain, and evolve the software product, which is divided into:
    \begin{itemize}
        \item \textit{Architectural-Documentation-Related Assets (AM2.1.1)} refer to the assets used to design, communicate, represent, maintain, and evolve the architectural representation of a software product. We have identified the following assets belonging to this type: \textit{Architectural Models} and \textit{Architectural Documentation}.
        \item \textit{Design-Related Assets (AM2.1.2)} refer to the assets that belong to the design that occurs during the development process. We have identified the following assets belonging to this type: \textit{Design Decisions Documentation} and \textit{System Designs}.
    \end{itemize}
    \item\textit{Product-Documentation-Related Assets (AM2.2)} refer to the assets that belong to the product documentation and the process of creating such documentation. We have identified the following assets belonging to this type: \textit{Documentation Automation Scripts} and \textit{Product Documentation}.
\end{itemize}

\begin{table*}[htpb]
\caption{Product-Representation-Related Assets’ Definitions.}
\begin{tabular}{p{2.5cm}p{1.2cm}p{12cm}}
\toprule
\textbf{Asset} & \textbf{AM type} & \textbf{Definition} \\
\midrule
Architectural Models & AM2.1.1 & Architecture Models are partial abstractions of systems, they capture different properties of the system~\cite{kruchten19954+}. ``Architecture modeling involves identifying the characteristics of the system and expressing it as models so that the system can be understood. Architecture models allow visualisation of information about the system represented by the model.''~\cite{kumar2014value}\\
Architectural Documentation & AM2.1.1 & Architectural Documentation are the representations of the decisions made to construct the architecture of the software~\cite{kruchten19954+}.\\
Design Decisions Documentation & AM2.1.2 & Design Decisions Documentation are the results of the design decisions that architects create and document during the architectural design process~\cite{bourque2014guide}.\\
System Designs & AM2.1.2 & System Designs are the processes of defining elements of a system. These  elements are specified in the requirements and are extracted to create modules, architecture, components and their interfaces and data for a system.\\
Documentation Automation Scripts & AM2.2 & Documentation Automation Scripts are the scripts that generate documentation based on the state of the source code.\\
Product Documentation & AM2.2 & Product Documentation are the operational guidelines (such as \textit{user manuals} and \textit{installation guides}) for when the product is in use.\\
\bottomrule
\end{tabular}
\label{tab:tab6_am2}
\end{table*}

\subsubsection{Development-Related Assets (AM3)}\label{sec:4_subsub_AM3}

\begin{figure*}[ht]
\centering
  \includegraphics[width=1\textwidth]{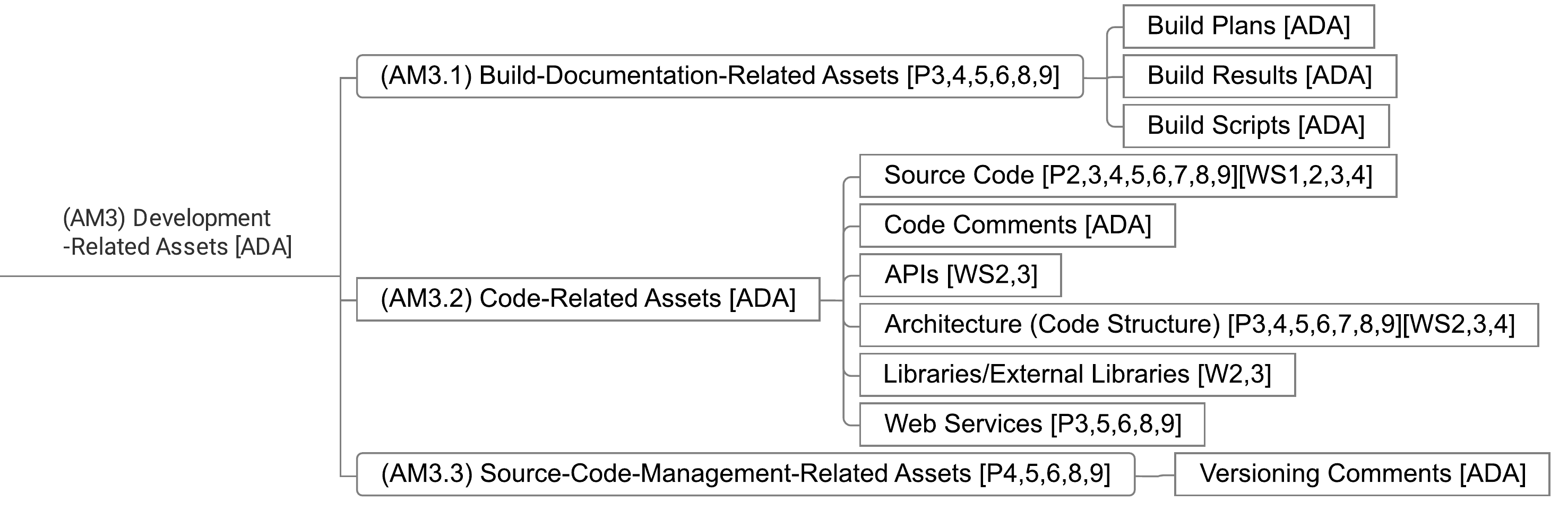}
\caption{Development-Related Assets Sub-tree.}
\label{fig:07_AM3}   
\end{figure*}

Development-Related Assets include the following three types of assets (see Figure~\ref{fig:07_AM3} with the sub-tree and Table~\ref{tab:tab7_am3} with definitions of each asset):

\begin{itemize}
    \item \textit{Build-Documentation-Related Assets (AM3.1)} refer to the assets related to the build system itself, the build environment, and the build process. We have identified the following assets belonging to this type: \textit{Build Plans}, \textit{Build Results}, and \textit{Build Scripts}.
    \item \textit{Code-Related Assets (AM3.2)} refer to the assets that are related to the source code. We have identified the following assets belonging to this type: \textit{Source Code}, \textit{Code Comments}, \textit{APIs}, \textit{Architecture (Code Structure)} ---i.e., a set of structures that can be used to reason about the system including the elements, relations among them, and their properties~\cite{bass2003software}---, and \textit{Libraries/External Libraries}.
    \item \textit{Source-Code-Management-Related Assets (AM3.3)} refer to the assets related to managing the source code, such as versioning and problems in code versioning and burndown charts. \textit{Versioning Comments} is an asset we identified belonging to this type.
\end{itemize}

\begin{table*}[htpb]
\caption{Development-Related Assets’ Definitions.}
\begin{tabular}{p{2.5cm}p{1.2cm}p{12cm}}
\toprule
\textbf{Asset} & \textbf{AM type} & \textbf{Definition} \\
\midrule
Build Plans & AM3.1 & Build Plans are the descriptions of how developers intend to build the software, i.e., by compilation of artefacts in a build chain, which will end in a running software.\\
Build Results & AM3.1 & Build Results are the results of the build process, including the comments, documentation, and other artefacts that are generated during the build process.This is seen as a persistent asset if it holds more data than just an automated ``throw away'' report, and/or if the asset is used for reference over time.\\
Build Scripts & AM3.1 & Build Scripts are the scripts that are used to run the build process.\\
Source Code & AM3.2 & Source Code is the collection of code written in a human-readable and comprehensible manner stored as plain text~\cite{kernighan1974programming}.\\
Code Comments & AM3.2 & Code Comments are the comments that developers integrate and write within the source code to clarify and describe certain parts of the code or its functionality~\cite{grubb2003software}.\\
APIs & AM3.2 & APIs (Application Program Interfaces) are the interfaces that are created to facilitate interaction of different components and modules. \\
Architecture (Code Structure) & AM3.2 & Architecture is the actual and fundamental relationships and structure of a software system and its source code~\cite{bass2003software}.\\
Libraries/ External Libraries & AM3.2 & Libraries/External Libraries are source code that belongs to the product but is not developed or maintained within the project, i.e., the developers. The software project depends on it and references the library.\\
Web Services & AM3.2 & Web Services are running services on devices handling requests coming from networks \\
Versioning Comments & AM3.3 & Versioning comments are the comments that developers submit to any version control application they use for the development. Such comments can later be extracted and viewed to identify the purpose of each event.\\
\bottomrule
\end{tabular}
\label{tab:tab7_am3}
\end{table*}

\subsubsection{Verification-and-Validation-Related Assets (AM4)}\label{sec:4_subsub_AM4}

\begin{figure*}[htpb]
\centering
  \includegraphics[width=1\textwidth]{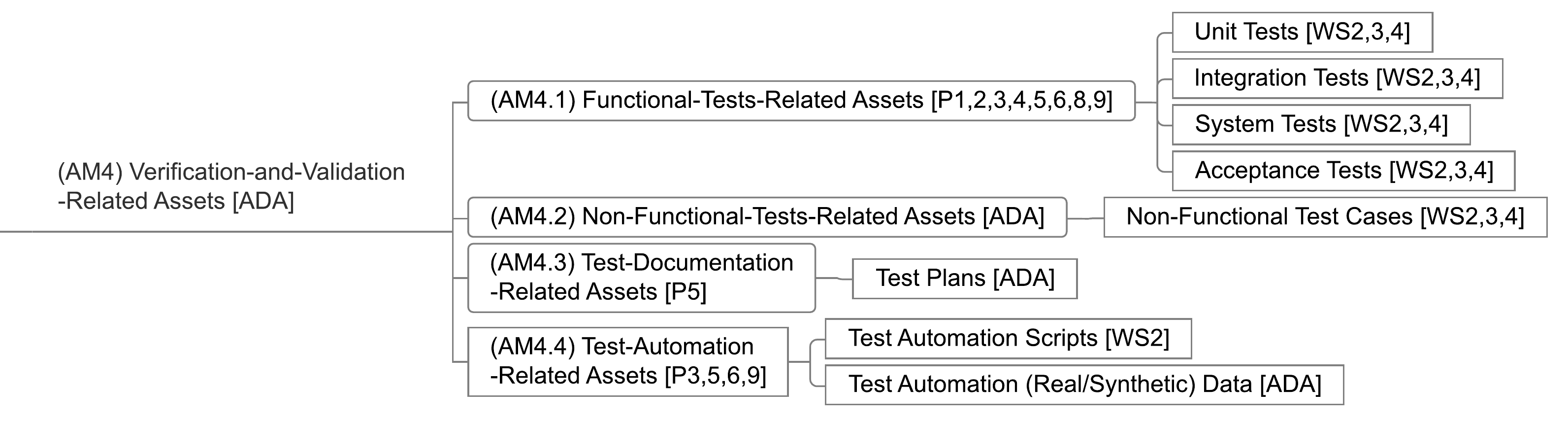}
\caption{Verification-and-Validation-Related Assets Sub-tree.}
\label{fig:08_AM4}   
\end{figure*}

Verification-and-Validation-Related Assets include the following four types of assets (see Figure~\ref{fig:08_AM4} with the sub-tree and Table~\ref{tab:tab8_am4} with definitions of each asset):

\begin{itemize}
    \item \textit{Functional-Tests-Related Assets (AM4.1)} refer to the assets related to testing the functionality of the system, its related features, and how they work together. We have identified the following assets belonging to this type: \textit{Unit Tests}, \textit{Integration Tests}, \textit{System Tests}, and \textit{Acceptance Tests}.
    \item \textit{Non-Functional-Test-Related Assets (AM4.2)} refer to the assets related to testing the quality attributes of the system and whether they satisfy the business goals and requirements. We have identified \textit{Non-Functional Test Cases} as an asset which belongs to this type.
    \item \textit{Test-Documentation-Related Assets (AM4.3)} refer to the assets related to documenting the testing process. \textit{Test Plans} is an asset we identified belonging to this type.
    \item \textit{Test-Automation-Related Assets (AM4.4)} refer to the assets that are utilised for automated testing of the system. We have identified the following assets belonging to this type: \textit{Test Automation Scripts} and \textit{Test Automation (Real/Synthetic) Data}.
\end{itemize}

\begin{table*}[htpb]
\caption{Verification-and-Validation-Related Assets Definitions.}
\begin{tabular}{p{2.5cm}p{1.2cm}p{12cm}}
\toprule
\textbf{Asset} & \textbf{AM type} & \textbf{Definition} \\
\midrule
Unit Tests & AM4.1 & Unit Tests are the tests written to examine the individual units of the code~\cite{hamill2004unit}. ``Unit tests generally focus on the program logic within a software component and on correct implementation of the component interface.''~\cite{berner2005}\\
Integration Tests & AM4.1 & Integration Tests are the tests written to examine the combined set of modules as a group~\cite{berner2005,ISO/IEC/IEEE2010}. \\
System Tests & AM4.1 & System Tests are the tests written to examine the system's compliance with the requirements.\\
Acceptance Tests & AM4.1 & Acceptance Tests are the tests conducted to examine and determine whether the requirements are met according to the specifications of the requirements.\\
Non-Functional Test Cases & AM4.2 & Non-Functional Test Cases are the tests that examine the quality of the system, i.e., non-functional aspects such as performance, availability, and scalability.\\
Test Plans & AM4.3 & Test Plans are the documents that describe the testing scope and test activities that will be performed on the system throughout the development lifecycle.\\
Test Automation Scripts & AM4.4 & Test Automation Scripts are the scripts that automate part of the testing process. More specifically, the scripts automate distinct testing activities or types of tests.\\
Test Automation (Real/ Synthetic) Data & AM4.4 & Test Automation (Real/Synthetic) Data is the generated data that are used by the automation scripts to test the system. \\
\bottomrule
\end{tabular}
\label{tab:tab8_am4}
\end{table*}

\subsubsection{Operations-Related Assets (AM5)}\label{sec:4_subsub_AM5}

Operation-Related Assets are all the assets created as the result of operational activities, extracted during the operational activities, or used during the operational activities, e.g., any data collected during the use of the product or service (see Figure~\ref{fig:09_AM5} with the sub-tree). The operations-related assets include \textit{Customer Data}, \textit{Application Data}, and \textit{Usage Data}. Table~\ref{tab:tab9_am5} provides definitions of each asset:

\begin{figure*}[htpb]
\centering
  \includegraphics[width=0.5\textwidth]{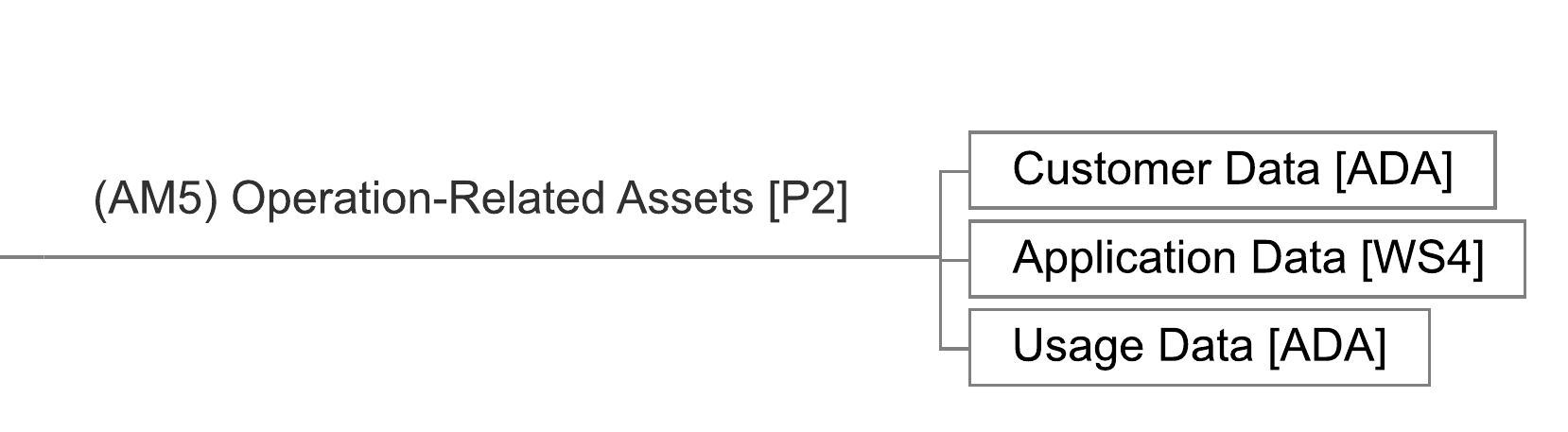}
\caption{Operation-Related Assets Sub-tree.}
\label{fig:09_AM5}   
\end{figure*}

\begin{table*}[htpb]
\caption{Operations-Related Assets’ Definitions.}
\begin{tabular}{p{2.5cm}p{1.2cm}p{12cm}}
\toprule
\textbf{Asset} & \textbf{AM type} & \textbf{Definition} \\
\midrule
Customer Data & AM5 & Customer Data is data that is collected from the customers (end users) of the software product such as user feedback.\\
Application Data & AM5 & Application Data is the data that is created, collected, used, and maintained while developing the software product such as system performance.\\
Usage Data & AM5 & Usage Data is the data that is collected while the software product is operational such as the data related to the performance of the system.\\
\bottomrule
\end{tabular}
\label{tab:tab9_am5}
\end{table*}

\subsubsection{Environment-and-Infrastructure-Related Assets (AM6)}\label{sec:4_subsub_AM6}

Environment-and-Infrastructure-Related Assets are all the assets used in the development environment or as an infrastructure for development during software development (see Figure~\ref{fig:10_AM6} with the sub-tree). The  environment-and-infrastructure-related assets include \textit{Deployment Infrastructure}, \textit{Tools}, and \textit{Tools Pipelines}. Table~\ref{tab:tab10_am6} presents environment-and- infrastructure -related assets definitions:

\begin{figure*}[htpb]
\centering
  \includegraphics[width=0.8\textwidth]{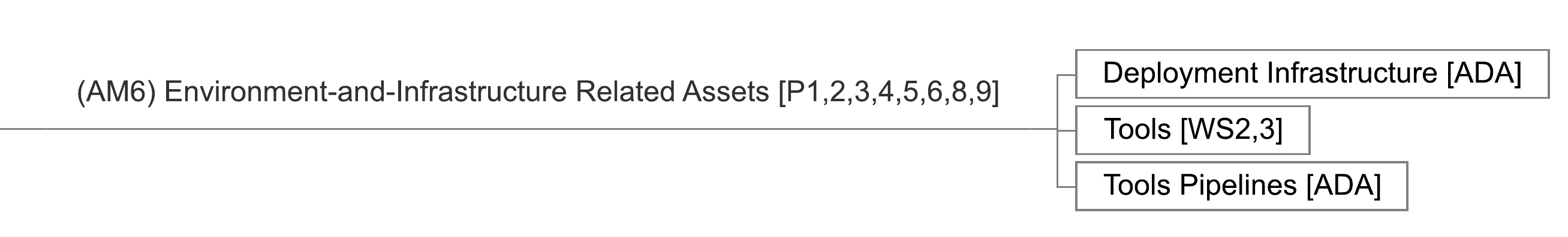}
\caption{Environment-and-Infrastructure-Related Assets Sub-tree.}
\label{fig:10_AM6}   
\end{figure*}

\begin{table*}[ht]
\caption{Environment-and-Infrastructure-Related Assets’ Definitions.}
\begin{tabular}{p{2.5cm}p{1.2cm}p{12cm}}
\toprule
\textbf{Asset} & \textbf{AM type} & \textbf{Definition} \\
\midrule
Deployment Infrastructure & AM6 & Deployment Infrastructure are all the steps, activities, tools, process descriptions, and processes that facilitate the deployment of a software-intensive product.\\
Tools & AM6 & Tools are any physical and virtual entities that are used for the development of a software product such as integrated development environments (IDE), version control systems, spreadsheets applications, compilers, and debuggers.\\
Tools Pipelines & AM6 & Tool Pipelines are automated processes and activities that facilitate and enable developers to reliably and efficiently compile, build, and deploy the software-intensive product.\\
\bottomrule
\end{tabular}
\label{tab:tab10_am6}
\end{table*}

\subsubsection{Development-Process/Ways-of-Working-Related Assets (AM7)}\label{sec:4_subsub_AM7}

\begin{figure*}[htpb]
\centering
  \includegraphics[width=1\textwidth]{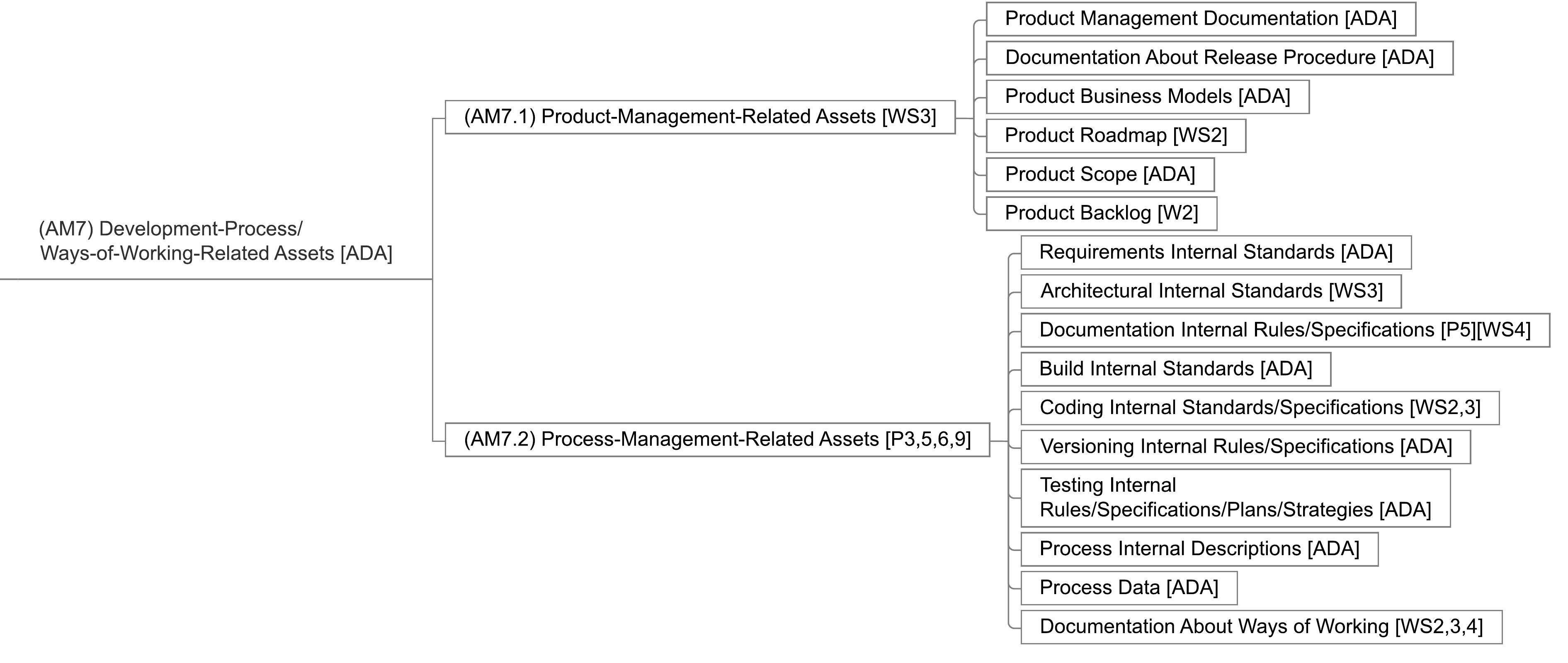}
\caption{Development-Process/Ways-of-Working-Related Assets Sub-tree.}
\label{fig:11_AM7}   
\end{figure*}

Development-Process / Ways-of-Working-Related Assets include the following three types of assets (see Figure~\ref{fig:11_AM7} with the sub-tree and Table~\ref{tab:tab11_am7} with definitions of each asset):

\begin{itemize}
    \item \textit{Product-Management-Related Assets (AM7.1)} refer to the assets related to the management of the product or service. These assets come from different stages, such as business justification, planning, development, verification, pricing, and product launching. We have identified the following assets belonging to this type: \textit{Product Management Documentation}, \textit{Documentation About Release Procedure}, \textit{Product Business Models}, \textit{Product Roadmap}, \textit{Product Scope}, and \textit{Product Backlog}.
    \item \textit{Process-Management-Related Assets (AM7.2)} refer to the assets related to managing the development process, including internal rules, plans, descriptions, specifications, strategies, and standards. We have identified the following assets belonging to this type: \textit{Requirements Internal Standards}, \textit{Architectural Internal Standards}, \textit{Documentation Internal Rules / Specifications}, \textit{Build Internal Standards}, \textit{Coding Internal Standards / Specifications}, \textit{Versioning Internal Rules / Specifications}, \textit{Testing Internal Rules / Specifications / Plans / Strategies}, \textit{Process Internal Descriptions}, \textit{Process Data}, and \textit{Documentation About Ways of Working}.
\end{itemize}

\begin{table*}[htpb]
\caption{Development-Process/Ways-of-Working-Related Assets’ Definitions.}
\begin{tabular}{p{2.5cm}p{1.2cm}p{12cm}}
\toprule
\textbf{Asset} & \textbf{AM type} & \textbf{Definition} \\
\midrule
Product Management Documentation & AM7.1 & Product Management Documentation is any documentation that is used to facilitate the management activities and processes during the product development.\\
Documentation About Release Procedure & AM7.1 & Documentation About Release Procedure is the description of the product release plan and the entities and activities associated with release.\\
Product Business Models & AM7.1 & Product Business Models are the descriptions of how the organisation creates value for the customers with the software-intensive product.\\
Product Roadmap & AM7.1 & Product Road Map is the abstract, high-level description of the evolution of the product during the development.\\
Product Scope & AM7.1 & Product Scope is the description of the characteristics, functionality, and features of the software-intensive product.\\
Product Backlog & AM7.1 & Product Backlog is any document that acts as a list where the features, change requests, bug fixes, and other similar activities are stored, listed, and prioritised.\\
Requirements Internal Standards & AM7.2 & Requirements Internal Standards are the specific rules that the company introduces and utilises internally for dealing with the requirements of the product.\\
Architectural Internal Standards & AM7.2 & Architectural Internal Standards are the specific rules that the development team introduces and utilises internally for designing, creating, and maintaining the architecture of the software-intensive product.\\
Documentation Internal Rules / Specifications & AM7.2 & Documentation Internal Rules/Specifications are the specific rules that the development team introduces and utilises internally for creating and maintaining the documentation.\\
Build Internal Standards & AM7.2 & Build Internal Standards are the specific rules that the development team introduces and utilises internally for the build activities.\\
Coding Internal Standards/Specifications& AM7.2 & Coding Internal Standards/Specifications are the rules that the development team introduces and utilises internally while developing the software-intensive product.\\
Versioning Internal Rules / Specifications & AM7.2 & Versioning Internal Rules/Specifications are the rules that the development team introduces and utilises internally for version control during the development of software-intensive products.\\
Testing Internal Rules / Specifications / Plans / Strategies & AM7.2 & Testing Internal Rules/Specifications/Plans/Strategies are the rules that the development team introduces and utilises internally for testing activities and procedures.\\
Process Internal Descriptions & AM7.2 & Process Internal Descriptions are the descriptions of the procedures and activities that the development team introduce and utilise during the development of software-intensive products.\\
Process Data & AM7.2 & Process Data is are the metrics and other information that concern the past and current status of the development process. Examples of such data are velocity, issues, bugs, backlog items, etc.\\
Documentation About Ways of Working & AM7.2 & Documentation About Ways of Working are the description of work plans and working patterns, i.e., how the organisation and the development team plan to create and release the software-intensive product.\\
\bottomrule
\end{tabular}
\label{tab:tab11_am7}
\end{table*}

\subsubsection{Organisation-Related Assets (AM8)}\label{sec:4_subsub_AM8}

Organisation-Related Assets are all the assets that represent organisations’ properties. The identified organisation-related assets include Organisational Structure, Organisational Strategy, and Business Models (see Figure~\ref{fig:12_AM8} with the sub-tree and Table~\ref{tab:tab12_am8} with the definition of each asset):

\begin{figure*}[htpb]
\centering
  \includegraphics[width=0.7\textwidth]{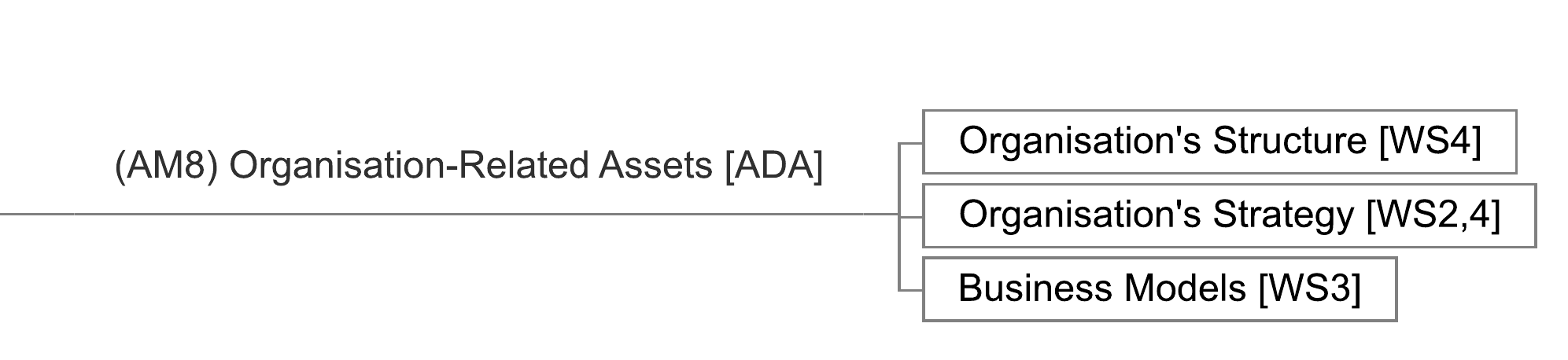}
\caption{Organisation-Related Assets Sub-tree.}
\label{fig:12_AM8}   
\end{figure*}

\begin{table*}[htpb]
\caption{Organisation-Related Assets’ Definitions.}
\begin{tabular}{p{2.5cm}p{1.2cm}p{12cm}}
\toprule
\textbf{Asset} & \textbf{AM type} & \textbf{Definition} \\
\midrule
Organisation's Structure & AM8 & Organisation’s Structure is the description of how the organisation directs the activities to achieve organisational goals.\\
Organisation's Strategy & AM8 & Organisation’s Strategy is the description of the plans that guide the organisation how to allocate its resources to support the development of the software-intensive product. \\
Business Models & AM8 & Business Models are the descriptions of how the organisation creates value with the software-intensive product for the organisation. \\
\bottomrule
\end{tabular}
\label{tab:tab12_am8}
\end{table*}

\subsection{Summary of Validation Workshops}\label{sec:validation_results}

In this section, we present the summary of the taxonomy validation workshop described in Section~\ref{sec:3_sub_internal_validation}.

The participants were asked to select the assets from the taxonomy that they were not aware of or assets that were new to them. Collectively the participants selected 33 assets. These assets and the number of times they were chosen are presented in Table~\ref{tab:vq1}. The participants were asked to select the assets that, based on their experience, should not be in the taxonomy. Overall, the participants selected 13 assets. These assets and the number of times there were chosen are presented in Table~\ref{tab:vq2}.

\begin{table}[htpb]
\caption{This table summarises the assets that the validation workshop participants were not aware of. The number represented the number of times the asset was selected by the participants.}
\scriptsize
\begin{tabular}{lll}
\toprule
& \textbf{Asset} & \textbf{Selected by} \\
\midrule
1 & \textbf{AM2.2} - Documentation Automation Scripts & 4\\
2 & \textbf{AM7.1} - Product Management Documentation & 3\\
3 & \textbf{AM1.2} - Quality Scenarios & 2\\
4 & \textbf{AM3.1} - Build Plans & 2\\
5 & \textbf{AM7.1} - Product Business Models & 2\\
6 & \textbf{AM7.2} - Requirements Internal Standards  & 2\\
7 & \textbf{AM7.2} - Architectural Internal Standards & 2\\
8 & \makecell[l]{\textbf{AM7.2} - Documentation Internal Rules\\ / Specifications} & 2\\
9 & \textbf{AM7.2} - Build Internal Standards & 2\\
10 & \makecell[l]{\textbf{AM7.2} - Versioning Internal Rules\\ / Specifications} & 2\\
11 & \makecell[l]{\textbf{AM7.2} - Testing Internal Rules\\ / Specifications / Plans / Strategies} & 2\\
12 & \textbf{AM7.2} - Process Internal Descriptions & 2\\
13 & \textbf{AM7.2} - Process Data & 2\\
14 & \textbf{AM1.}1 Feature-Related Backlog Items & 1\\
15 & \textbf{AM1.2} - User Experience Requirements & 1\\
16 & \textbf{AM2.1.1} - Architectural Models & 1\\
17 & \textbf{AM3.1} - Build Results  & 1\\
18 & \textbf{AM3.2} - Architecture (Code Structure) & 1\\
19 & \textbf{AM3.2} - Libraries / External Libraries & 1\\
20 & \textbf{AM3.2} - Web Services & 1\\
21 & \textbf{AM3.3} - Versioning Comments & 1\\
22 & \textbf{AM4.1} - Acceptance Tests & 1\\
23 & \makecell[l]{\textbf{AM4.4} - Test Automation (Real\\ / Synthetic) Data} & 1\\
24 & \textbf{AM5} - Usage Data & 1\\
25 & \textbf{AM6} - Deployment Infrastructure & 1\\
26 & \textbf{AM6} - Tools & 1\\
27 & \makecell[l]{\textbf{AM7.1} - Documentation About Release\\ Procedure} & 1\\
28 & \textbf{AM7.1} - Product Scope & 1\\
29 & \textbf{AM7.1} - Product Backlog & 1\\
30 & \makecell[l]{\textbf{AM7.2} - Coding Internal Standards\\ / Specifications} & 1\\
31 & \makecell[l]{\textbf{AM7.2} - Documentation About\\ Ways of Working} & 1\\
32 & \textbf{AM8} - Organisation's Structure & 1\\
33 & \textbf{AM8} - Organisation's Strategy & 1\\
\bottomrule
\end{tabular}
\label{tab:vq1}
\end{table}
\begin{table}[htpb]
\caption{This table summarises the assets that the validation workshop participants, based on their experience, deemed not needed to be included in the taxonomy. The number represented the number of times the asset was selected by the participants.}
\scriptsize
\begin{tabular}{lll}
\toprule
& \textbf{Asset} & \textbf{Selected by} \\
\midrule
1 & \textbf{AM1.2} - Quality Scenarios & 1\\
2 & \textbf{AM3.1} - Build Plans & 1\\
3 & \textbf{AM3.2} - Libraries / External Libraries & 1\\
4 & \textbf{AM4.1} - Unit Tests & 1\\
5 & \textbf{AM5} - Usage Data & 1\\
6 & \textbf{AM7.1} - Product Management Documentation  & 1\\
7 & \textbf{AM7.2} - Requirements Internal Standards & 1\\
8 & \textbf{AM7.2} - Architectural Internal Standards & 1\\
9 & \makecell[l]{\textbf{AM7.2} - Documentation Internal Rules\\ / Specifications} & 1\\
10 & \textbf{AM7.2} - Build Internal Standards & 1\\
11 & \makecell[l]{\textbf{AM7.2} - Versioning Internal Rules\\ / Specifications} & 1\\
12 & \makecell[l]{\textbf{AM7.2} - Testing Internal Rules\\ / Specifications / Plans / Strategies} & 1\\
13 & \textbf{AM8} - Business Models & 1\\
\bottomrule
\end{tabular}
\label{tab:vq2}
\end{table}

In an open question, the participants were asked to write down the assets that were missing from the taxonomy. \textit{Git Comments}, \textit{Code Review Data}, \textit{Contract Documentation}, and \textit{Communication Channels} were mentioned in the questionnaire. Finally, we asked the participants to prioritise the top five assets based on their experience. The answers to the last question are summarised in Table~\ref{tab:vt}.

The utility of the taxonomy is demonstrated through the use of taxonomy by practitioners and classification of existing knowledge~\cite{britto2016extended}. The participants in the validation workshops were able to use the taxonomy to identify and classify assets.

There are no existing taxonomies to compare the classification schemes; therefore, the taxonomy benchmarking was done with experts' knowledge. The validation workshop participants did not make any suggestions regarding the structure of the taxonomy. All participants agreed that the schema and categories (i.e., types of assets) were adequate.

\begin{table}[htpb]
\caption{This table summarises the assets that the validation workshop participants, based on their experience, selected as the top five assets. The number represented the number of times the asset was selected by the participants.}
\scriptsize
\begin{tabular}{ll}
\toprule
\textbf{Asset} & \textbf{Selected by} \\
\midrule
\textbf{AM1} - Product Requirements Related Assets & 4\\
\textbf{AM1.1} - Functional Requirements Related Assets & 1\\
\textbf{AM1.3} - Product Modification Related Assets & 1\\
Use Cases & 1\\
System Requirements & 1\\
\textbf{AM2} - Product Representation Related Assets  & 1\\
\textbf{AM2.1.2} - Design Related Assets & 2\\
Product Documentation & 1\\
\makecell[l]{\textbf{AM3} - Development Related Assets} & 3\\
\textbf{AM3.2} - Code Related Assets & 1\\
Source Code & 1\\
Architecture & 1\\
\makecell[l]{\textbf{AM4} - Verification and Validation Related Assets} & 2\\
\makecell[l]{\textbf{AM4.1} - Functional Tests Related Assets} & 1\\
\textbf{AM5} - Operation Related Assets & 1\\
\makecell[l]{\textbf{AM6} - Environment and Infrastructure\\Related Assets} & 3\\
\makecell[l]{\textbf{AM7} - Development Process / Ways\\of Working Related Assets} & 3\\
\textbf{AM7.1} - Product Management Related Assets & 1\\
\textbf{AM8} - Organisation Related Assets & 1\\
\bottomrule
\end{tabular}
\label{tab:vt}
\end{table}

\section{Discussion}\label{sec:discussion}
This section discusses our findings in light of the research question, followed by the general lessons learned and implications.
 
\subsection{Principal Findings}\label{sec:5_sub_findings}
 
$RQ:$ \textit{What assets are managed by organisations during the inception, planning, development, evolution, and maintenance of software-intensive products or services?}
 
In Section~\ref{sec:4_AM_taxonomy} we have presented a taxonomy of assets, which includes eight major types of assets \textit{AM1} to \textit{AM8}. Although the taxonomy is orthogonal by design (i.e., an asset or a type of asset can only be classified as a member of one type of assets), assets and types of assets are not isolated, i.e., some assets and types of assets are interrelated. For example, \textit{architectural documentation} is directly related to \textit{architecture} since architectural documentation is a representation of the architecture of the system.

During internal workshops and the taxonomy creation procedure, presented in Section~\ref{sec:3_sub_taxonomy_creation}, we identified several meta-characteristics of assets. These meta-characteristics are:

\begin{itemize}
    \item \textbf{Easier to contextualise}: It is easier for the stakeholders to identify such assets in the software product context. For example, the data that the company acquires from the operation of the product, i.e., \textit{Application Data}, can be used as input to improve the product.
    \item \textbf{More tangible}: Some assets are more prominent in industry and have been studied and discussed more deeply before, and therefore the asset is not alien anymore. For example, every software company, one way or another, has \textit{Code}, in one form or another, or a \textit{Product Backlog} with specific characteristics which is familiar to all people involved in the development of the software-intensive products or services.
    \item \textbf{Easier to measure}: There are already existing metrics used to measure the state of such assets. For example, there are many metrics available to measure \textit{Source Code}, such as LOC and Cyclomatic Complexity.
    \item \textbf{Used universally}: The assets that are defined in the same way across different organisations and academia, meaning that they are not organisation-specific. For example, the \textit{software's architecture (Code Structure)} is a universal and inherent aspect of any software-intensive product or service.
\end{itemize}

Out of the eight major types of assets, two types have been studied more extensively, namely \textit{Development-Related} Assets and \textit{Product-Representation-Related} Assets. These results are not new and have been highlighted in previous studies such as~\cite{avgeriou2016managing}, i.e., prior studies on TD focus on source-code-related assets. These types of assets are easier to study due to the abundance of metrics and evaluation methods and, therefore, have been studied in many research articles. The reason behind this might be that: 
\begin{itemize}
    \item The TD metaphor was initially introduced in the context of prevalent assets~\cite{avgeriou2016managing}. Therefore the researchers have spent more time investigating and exploring this specific phenomenon. For example, many papers investigate a software product's architecture, exploring different ways of evaluating architecture using different tools and measurements.
    \item These types of assets are easier to contextualise in the TD metaphor, i.e., identifying such assets and how they can be subject to incur debt. For example, the concept of code smells is easier to grasp since it is a more tangible artefact. It is simple to define how the software product can incur debt if the code does not align with certain ``gold standards''; i.e., it is smelly.
\end{itemize}

The rest of the types of assets have not received extensive time to be explored. The reason behind this might be that:
\begin{itemize}
    \item These types of assets were added later as ``types of technical debt,'' such as Requirements Debt~\cite{Ernst2012,Lenarduzzi2019} and Process Debt~\cite{Martini2019}. The TD metaphor was not initially used to deal with these types of assets~\cite{avgeriou2016managing}. These types of TD were introduced in an effort to extend the metaphor and, therefore, have not been investigated thoroughly.
    \item Unlike the other types (i.e., \textit{Development-Related} and \textit{Product-Representation-Related} Assets), it is harder to identify and/or define how and to what extent one can incur debt in software products. For example, incurring Documentation Debt might differ in different companies and development teams.
\end{itemize}

We have seen that the existing literature on TD classifies various TD types and presents ontologies on the topic. These classifications have evolved since the introduction of the extended TD metaphor. We observe that the relevant asset categories we have extracted from industrial insights can be mapped to the classifications provided in TD literature. We observe that:
\begin{itemize}
    \item Some existing TD types and categories, such as code that are well-defined and well-recognised, fit into similar categories as in the presented taxonomy. 
    \item Some types of assets that are relevant to the industry have been understudied or not even studied at all. There is room for extending the research in such areas~\cite{rios2018tertiary}, e.g., \textit{Operations-Related Assets (AM5)} and \textit{Environment-and-Infrastructure-Related Assets (AM6)} (see Section~\ref{sec:4_sub_results_literature_review}).
    \item By creating the taxonomy, we highlight both the areas of interest and the gaps in research. Therefore, identifying the areas in the software engineering body of knowledge that need to be investigated and the areas that need to evolve according to the current interest.
\end{itemize}

Finally, our taxonomy of assets has an innate relationship with the TD research and the taxonomies, ontologies, secondary, and tertiary studies in the TD topic. The taxonomy of assets is created using empirical evidence from the industry and peer-reviewed articles. The taxonomy of assets provides actual tangible assets, whereas other studies mainly present areas where assets belong to. Furthermore, we present our results from the industrial perspective and how the TD metaphor is not enough to deal with assets' problems, i.e., TD's main focus is sub-optimal solutions. In contrast, asset degradation goes beyond TD and considers different types of degradation, including the ripple effect and chain reactions of degradation that stems from the relations between assets. From what we gathered, the industry does not care about TD in isolation. TD is a part of a bigger problem, i.e., managing asset degradation when companies deal with software-intensive products or services~\cite{zabardast2022assets}. And there is a need to consider a more holistic view to be able to address the asset degradation problem.

\subsection{Lessons Learned}\label{sec:5_sub_lessons}
 
This section presents the lessons learned from running the industrial focus groups, synthesising the findings, and creating the taxonomy. The importance of source-code-related assets is undeniable (i.e., assets in \textit{AM1}, \textit{AM2}, \textit{AM3}, and \textit{AM4}). However, we observe that the social and organisation aspect of the development is very important to the industry, through these aspects have not received as much attention in the TD area~\cite{avgeriou2016managing,rios2018tertiary}, although the TD community has already identified them as a research area that deserves more attention~\cite{Martini2019}. Taking a look at some statements from participants in the industrial focus groups highlights this fact. Examples of such statements are:
\begin{itemize}
    \item \textit{``There are many people who work in the same area in the same code base. Creates conflicts and slow releases.''}
    \item \textit{``The problem is the delta operation, and the plan is at such a high level that it is impossible to understand. Too abstract.''}
    \item \textit{``... training the teams in what is considered best practices improves team cohesion and eases collaboration.''}
    \item \textit{``[There is] no holistic platform strategy (Conway's law).''}
\end{itemize}

The large-scale software projects developed in large organisations are highly coupled with the social and organisation aspect of work. The prevalence of assets related to the social and organisation aspect of development, e.g., \textit{Business Models} and \textit{Product Management Documentation}, indicates the necessity to characterise and standardise such assets, how they are perceived, and how they are measured and monitored.

While creating the taxonomy, we observed that assets do not exist in isolation, i.e., they are entities with characteristics and properties that exist in a software development environment. In the following, we will discuss the assets with similar characteristics and properties and assets that have implicit relations with each other.\\

\noindent \textbf{Assets that have similar characteristics and properties.} For example, \textit{Unit Tests} have similar characteristics and properties as \textit{Source Code}, i.e., unit tests are code, and therefore, their value degradation can have analogous connotations. This means that there are possibilities to evaluate such assets' degradation with similar characteristics and properties using similar metrics. Still, the degradation of one asset (e.g., Source Code) might impact or even imply the degradation of the other asset (e.g., Unit Tests)~\cite{Alegroth2017}. Therefore, such coupling and relations of the assets should be considered when analysing and managing such assets.\\

\noindent \textbf{Assets that have implicit relations between each other.} Implicit relations between assets can arise from their inherent coupling properties. Different assets related to certain aspects of the product will have implicit relations that are not visible in the taxonomy as presented now. For example, \textit{Architectural Models} and \textit{Architecture (Code Structure)} have an inherent relationship. Architectural Models should be the representation of the architecture of the system, i.e., the code structure. Therefore, similar to the previous point, the value degradation of the assets with such implicit relations can have analogous connotations. Their degradation might impact the degradation of the other related assets. For example, the degradation of any of the functional-requirements-related assets will eventually be reflected in the degradation of functional-test-related assets.

\subsection{Contributions}\label{sec:5_sub_contributions}
 
The contribution of this work is the following: 
\begin{itemize}
    \item Providing common terminology and taxonomy for assets that are utilised during software development and;
    \item Providing a mapping over the assets and the input used to create the taxonomy, i.e., input from the literature and input from the industry
\end{itemize}

One contribution of the taxonomy is that it is a guideline for future research by providing a map of different types of assets. The map illustrates the different areas defined and studied and those that lack a shared understanding or are under-explored. Therefore, the taxonomy provides a summary of the body of knowledge by linking empirical studies with industrial insights gathered through the industrial focus groups. Providing a common taxonomy and vocabulary:
\begin{itemize}
    \item Makes it easier for the different communities to communicate the knowledge.
    \item Creates the opportunity to find and build upon previous work.
    \item Helps to identify the gaps by linking the empirical studies to the taxonomy.
    \item Highlights potential areas of interest.
    \item Makes it possible to build and add to the taxonomy (new assets, details) as knowledge is changed over time by researchers in the field.
    \item For practitioners, the evolving taxonomy can be used as a map to identify assets and the degradation that can impact other assets (ripple and chain effects that can spread the degradation to other assets like degradation of code causing degradation on test-code or vice-versa~\cite{Alegroth2017}), where such implicit degradation are not immediately detectable.
\end{itemize}

Finally, it might help large organisations to deal with developing software-intensive products or services that rely on external resources to help them achieve the business goals of their products. A major external contributor to new knowledge that can help practitioners in the industry is research findings. Therefore, understanding and applying the research findings is crucial for them. Having a taxonomy of assets summarising the state-of-the-art and state-of-practice body of knowledge for the assets utilised for developing software-intensive products or services is useful. Practitioners can refer to the taxonomy systematically built with the accumulated knowledge of academia and other practitioners to extract what they need in specific domains.

We believe that our findings to collect and organise the different assets and terminologies used to describe the assets will help practitioners be more aware of each type of asset and how they are managed in the context.

\subsection{Limitations and Threats to Validity}\label{sec:5_sub_limitations_threats}
 
In this section, we cover the limitations of our work and how they might affect the results. The taxonomy is created based on the data extracted from the literature review, the field study, and academic expert knowledge. We combined the inputs from the mentioned sources to create the taxonomy. We designed the taxonomy to be extendable with new data identified by us and others in future studies as software engineering areas evolve (See Section~\ref{sec:3_sub_taxonomy_creation}). We encourage researchers and practitioners to consider the taxonomy within their organisation and identify the potential new types of assets and assets that can complement the asset management taxonomy's representativeness.
 
In the rest of this section, we cover the threats to construct, internal, and external validity as suggested by Runeson and H{\"o}st~\cite{runeson2009guidelines} and Runeson et al.~\cite{runeson2012case}.\\

\noindent \textit{Construct validity} reflects the operational measurements and how the study represents what is being investigated. We are aware that the literature review is conducted in a limited area, i.e., TD. We chose the TD topic for the reasons mentioned in Section~\ref{sec:3_sub_literature_review}. We acknowledge that the performed snowballing and limiting the literature review to a specific topic might affect the construct validity of this work since we could not cover all the topics and the articles related to those topics. 
We acknowledge that the participants of the focus groups do not include all the relevant stakeholders in organisations. We tried mitigating this threat by involving participants with different roles and varying expertise from the companies. We are also aware that the participants' statements in the focus groups can be interpreted differently by the researcher and the participants. We mitigate this threat in three ways. First, by sending the summary, with the transcription of their statements and our own notes, back to the participants, asking for their feedback; second, by having two researchers code the raw data independently; and third, by choosing to code the data using the \textit{in vivo} coding method, the qualitative analysis prioritises the participants' opinions.\\

\noindent \textit{External validity} refers to the generalisability of the results and whether the results of a particular study can hold in other cases. We acknowledge and understand that the results are not comprehensive and might not be generalisable. The created taxonomy is based on the collected data and is extendable. We have provided a systematic way of extending the taxonomy, i.e., the meta-model. Finally, other threats that can affect the study's external validity are the number of involved companies, the country where the companies (investigated sites) are located, i.e., Sweden, and the involvement of all the roles in these organisations.\\ 

\noindent \textit{Reliability} refers to the extent that the data and analysis are dependent on the researchers. When conducting qualitative studies, the goal is to provide results that are consistent with the collected data~\cite{merriam2015qualitative}. We have tried to mitigate this threat, i.e., consistency of the results. Firstly, by rigorously documenting and following the procedures of the focus groups to collect consistent data~\cite{yin2009robert}. And secondly, by relying on consistency during the analysis, i.e., blind labelling of the data by multiple researchers and peer reviewing the labels.

\section{Conclusions and Future Work}\label{sec:conclusions_futurework}

This paper presents a taxonomy for classifying assets with inherent value for an organisation subject to degradation. These assets are used during the development of software-intensive products or services. The creation of the taxonomy of assets attempts to provide an overarching perspective on various assets for researchers and practitioners. The taxonomy allows us to characterise and organise the body of knowledge by providing a common vocabulary of and for assets. 

Eight major types of assets are introduced in the taxonomy: assets related to \textit{Product Requirements}, \textit{Product Representation}, \textit{Development}, \textit{Verification and Validation}, \textit{Operations}, \textit{Environment and Infrastructure}, \textit{Development Process/ Ways-of-Working}, and \textit{Organisation}.

The taxonomy could be used for:
\begin{itemize}
    \item Identify the gaps in research by providing the points of interest from practitioners' perspectives.
    \item Identify the state-of-the-art research for individual assets and their properties for practitioners.
    \item Communicate and spread the body of knowledge.
\end{itemize}

The taxonomy helps draw out the assets with similar characteristics and implicit relations among them. Most of such similarities of characteristics, properties, and relations are not immediately visible when considering the assets from only one perspective. Taking a more abstract and high-level look at the assets involved in the development of software-intensive products or services can help facilitate the management activities and the overall development process.

The dimensions provided by our taxonomy are not exhaustive, nor are the assets we identified. Therefore, we intend to conduct further investigation to complement the taxonomy by incorporating the new knowledge. Furthermore, we would like to study the relationship between assets, how the degradation of an asset can have a ripple effect and chain reactions, causing the degradation of other assets, and how the degradation impacts the development process. Lastly, we intend to investigate the individual properties of assets to identify the metrics used for measuring assets, their value, and their degradation (or lack thereof). 

Besides, future and ongoing work will use the taxonomy as a base for further studies and exploration of assets, their characteristics, and the concepts of value, degradation and its different types.

\appendix
\section{Appendix: The Asset Management Taxonomy}\label{appendix}

\begin{figure*}[h]
\centering
  \includegraphics[width=0.95\textwidth]{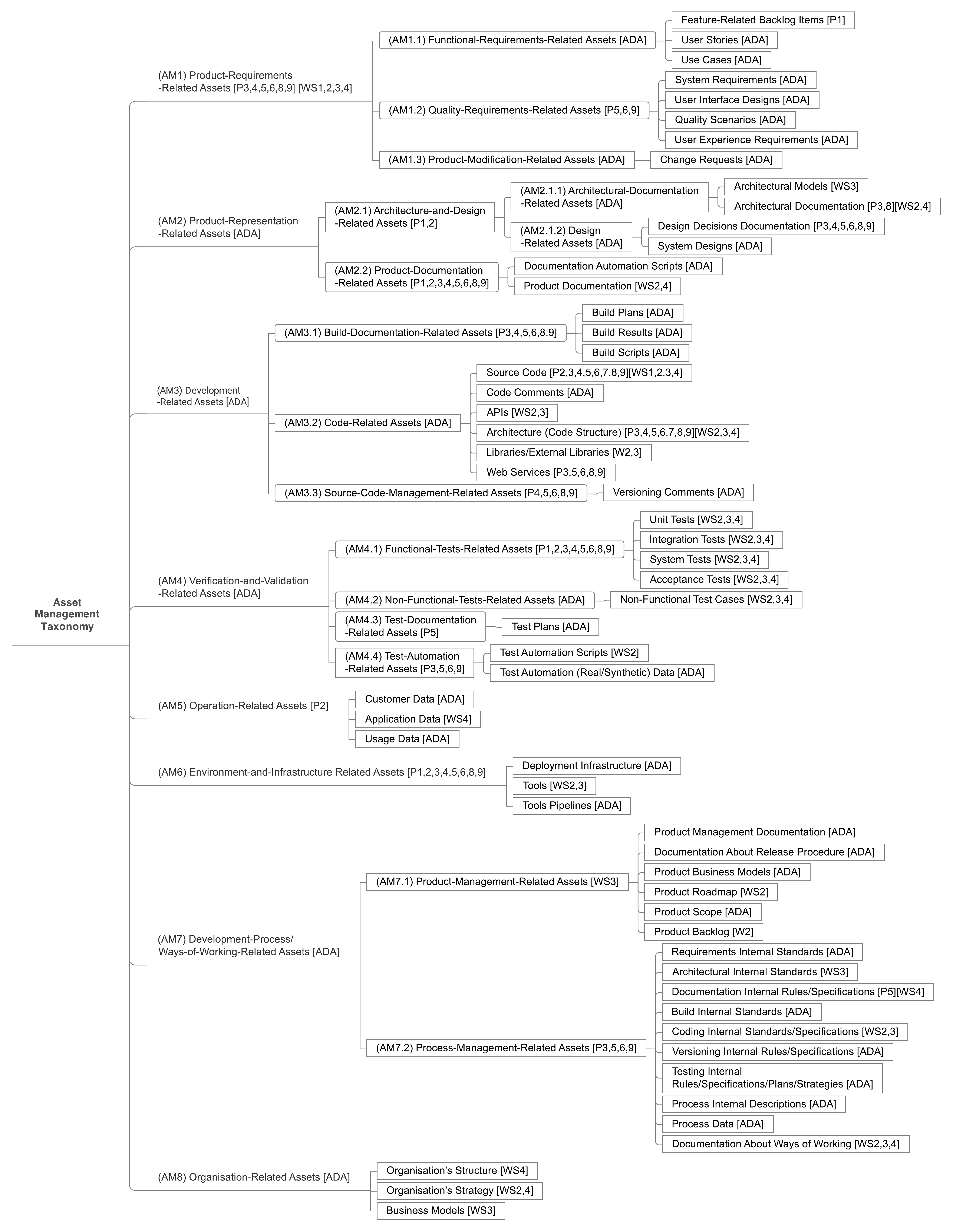}
\caption{The asset management taxonomy.}
\label{fig:AMT_full}   
\end{figure*}

The full tree of the asset management taxonomy is presented in Figure~\ref{fig:AMT_full}.

\printcredits

\bibliographystyle{cas-model2-names}

\bibliography{references}



\end{document}